\documentclass{article}

\usepackage{PRIMEarxiv}

\usepackage[utf8]{inputenc} % allow utf-8 input
\usepackage[T1]{fontenc}    % use 8-bit T1 fonts
\usepackage{hyperref}       % hyperlinks
\usepackage{url}            % simple URL typesetting
\usepackage{booktabs}       % professional-quality tables
\usepackage{amsfonts}       % blackboard math symbols
\usepackage{nicefrac}       % compact symbols for 1/2, etc.
\usepackage{microtype}      % microtypography

\usepackage{fancyhdr}       % header
\usepackage{graphicx}       % graphics
\graphicspath{{media/}}     % organize your images and other figures under media/ folder
\usepackage{float}

\usepackage{chemformula}
\usepackage{siunitx}
\usepackage{gensymb}
\usepackage{soul}
\usepackage[nomain, acronym]{glossaries}

\newacronym{ANN}{ANN}{Artificial Neural Network}
\newacronym{FDTD}{FDTD}{Finite-Difference Time-Domain}
\newacronym{PML}{PML}{Perfectly Matched Layer}
\newacronym{MLP}{MLP}{Multilayer Perceptron}
\newacronym{GELU}{GELU}{Gaussian Error Linear Unit}
\newacronym{MSE}{MSE}{Mean Square Error}
\newacronym{FWHM}{FWHM}{Full Width at Half Maximum}
\newacronym{FEM}{FEM}{Finite Element Method}
\newacronym{DGTD}{DGTD}{Discontinuous Galerkin Time Domain}
\newacronym{ReLU}{ReLU}{Rectified Linear Unit}
\newacronym{MVNN}{MVANN}{Multi-Valued Artificial Neural Network}
\newacronym{IR}{IR}{Infrared}
\newacronym{FB}{FB}{Fabry-Perot}
\newacronym{UV}{UV}{Ultraviolet}
\newacronym{NIL}{NIL}{Nanoimprint Lithography}
\newacronym{PINN}{PINN}{Physics-Informed Neural Network}
\newacronym{FFT}{FFT}{Fast Fourier Transform}
\newacronym{DL}{DL}{Deep Learning}
\newacronym{GA}{GA}{Genetic Algorithm}
\newacronym{PSO}{PSO}{Particle Swarm Optmization}
\newacronym{ACO}{ACO}{Ant Colony Optimization}
\newacronym{GP}{GP}{Gaussian Process}
\newacronym{GPU}{GPU}{Graphics Processing Unit}
\newacronym{CPU}{CPU}{Central Processing Unit}
\newacronym{BP}{BP}{Back-Propagation}
\newacronym{BN}{BN}{Batch Normalization}

\title{Back-Propagation Optimization and Multi-Valued Artificial Neural Networks for Highly Vivid Structural Color Filter Metasurfaces
}

\author{
  Arthur Clini de Souza \\
  Université Côte d’Azur, Inria, CNRS, LJAD, 06902 Sophia Antipolis Cedex, France\\
  Laboratory of Applied and Computational Electromagnetism (LEMAC),\\
  School of Electrical and Computer Engineering (FEEC),\\
  University of Campinas (UNICAMP), Campinas, São Paulo, Brazil\\
  Solnil, 95 Rue de la République, 13002 Marseille, France\\
   \And
  Stéphane Lanteri \\
  Université Côte d’Azur, Inria, CNRS, LJAD,\\
  06902 Sophia Antipolis Cedex, France\\
  \And
  Hugo Enrique Hernandez-Figueroa\\
  Laboratory of Applied and Computational Electromagnetism (LEMAC),\\
  School of Electrical and Computer Engineering (FEEC),\\
  University of Campinas (UNICAMP), Campinas, São Paulo, Brazil\\
  \And
  Marco Abbarchi\\
  Solnil, 95 Rue de la République,\\
  13002 Marseille, France\\
  Université Aix Marseille, CNRS,\\
  Université de Toulon, IM2NP, UMR 7334,\\
  F-13397 Marseille, France\\
  \And
  David Grosso\\
  Solnil, 95 Rue de la République,\\
  13002 Marseille, France\\
  Université Aix Marseille, CNRS,\\
  Université de Toulon, IM2NP, UMR 7334,\\
  F-13397 Marseille, France\\
  \And
  Badre Kerzabi\\
  Solnil, 95 Rue de la République,\\
  13002 Marseille, France\\
  \And
  Mahmoud Elsawy$^*$\\
  Université Côte d’Azur, Inria, CNRS, LJAD,\\
  06902 Sophia Antipolis Cedex, France\\
  \texttt{mahmoud.elsawy@inria.fr}\\
}

\begin{document}
\maketitle

\begin{abstract}
We introduce a novel technique for designing color filter metasurfaces  using a data-driven approach based on deep learning. Our innovative approach employs inverse design principles to identify highly efficient designs that outperform all the configurations in the dataset, which consists of 585 distinct geometries solely. By combining Multi-Valued Artificial Neural Networks and back-propagation optimization, we overcome the limitations of previous approaches, such as poor performance due to extrapolation and undesired local minima. Consequently, we successfully create reliable and highly efficient configurations for metasurface color filters capable of producing exceptionally vivid colors that go beyond the sRGB gamut. Furthermore, our deep learning technique can be extended to design various pixellated metasurface configurations with different functionalities.
\end{abstract}
\section*{Introduction}
Optical color filters are structures or materials designed to discriminate and manipulate distinct light wavelengths through the selective transmission or reflection of particular colors while simultaneously absorbing or attenuating undesired colors \cite{bohren2008absorption, kerker2016scattering}. 
Conventional color filters rely on the manipulation of chemical composition to achieve the desired optical properties, which can lead to issues such as absorption losses, thermal effects, and alterations in chemical characteristics \cite{brown1993predicting}. An alternative approach involves the utilisation of structural color filters, offering distinct advantages and applications in diverse fields such as photorealistic color printing, color holography, anti-counterfeiting devices, and much more \cite{kim2009structural, chen2014high, hu2012photonic}.

Metasurfaces have emerged as a promising platform for structural color filters \cite{ding2017gradient,khaidarov2022large}, owing to its peculiar capability of controlling all the light properties at the nanoscale, enabling a plethora of applications \cite{kuznetsov2016optically, zheng2015metasurface, arbabi2018full, tittl2018imaging}. Dielectric metasurfaces play a crucial role in color filter applications, especially within the visible spectrum range where the plasmonic conterpart based on metals is less performing owing to intrinsic optical losses. The limited  losses of dielectrics (e.g. \ch{Si3N4}, \ch{GaN}, \ch{TiO2}, \ch{ZrO2}, \ch{HfO2}) make them highly desirable for designing efficient devices with sharp resonance responses \cite{yang2014all, arbabi2015dielectric, shalaev2015high, lin2014dielectric}.
Resonant dielectric metasurfaces achieve precise control over the phase of reflected and transmitted light by leveraging various resonant phenomena (e.g. Mie resonances) \cite{decker2015high, yu2015high}. Through meticulous engineering of the resonators, selective interaction with different wavelengths is enabled, leading to efficient and vivid color filters. Such kind of metasurfaces offer exceptional phase control, high-quality factors, and sharp resonances, resulting in enhanced color purity and spectral selectivity \cite{um2023dynamic,lin2023efficient,shang2023polarization}. Yet, the design of an ideal color filter demands capability to selectively filter all colors across the optical spectrum. In other words, at each desired wavelength, it is crucial to eliminate any background resonances in order to achieve a pure color response characterized by sharp reflection or transmission amplitudes. Given the fabrication constraints, finding the appropriate resonator shape to achieve a desired response, is a challenging task that has garnered significant attention in the research community. Numerous studies explored this area, employing sophisticated optimization algorithms including advanced \gls{DL} approaches to tackle the inherent complexity of the problem \cite{shang2023polarization,ma2019probabilistic,peurifoy2018nanophotonic,asano2018optimization,jiang2019global,wen2020robust,lin2023efficient, wiecha2019deep, wiecha2021deep, hemmatyar2019full}. 
However, relying on classical optimization approaches requires several costly simulations when optimizing various color targets simultaneously \cite{elsawy2019global, xu2021efficient, jafar2018adaptive}. A viable solution for the design of vivid metasurface color filters is one-shot optimization using \gls{ANN}. However, it is not straightforward owing to the presence of several designs with similar optical response whereas regular \gls{ANN} has only a single output \cite{zhang2018multivalued}.

%%%%%%%%%%
Here we present a novel data-driven methodology for efficiently designing fabrication constrained color filter metasurfaces. Our approach combines the ability to find suitable designs of \glspl{MVNN} with the solution refinement of back-propagation optimization. Thereby overcoming the fundamental limitations of relying on either the latter, which will lead to an undesirable local minima, or solely on a \gls{MVNN} leading to poor performance associated with extrapolation  \cite{zhang2018multivalued, peurifoy2018nanophotonic, asano2018optimization}. In our case, relying solely on \textcolor{black}{585} simulations and by varying four parameters allows for the optimization of a continuous spectrum of objectives and the identification of highly vivid  metasurface color filters. The optimized geometries exhibit a single sharp resonance response representing all the colors across the visible regime. To the best of our knowledge, our research outcomes exceed the previous findings documented in the literature, positioning our color filter as the foremost advancement in terms of vividness and overall performance \cite{liu2022transfer, panda2020transmission, wang2022tunable, khaidarov2022large, hemmatyar2019full}. 

%%%%%%%%%%%%%%%%%%%%%%%%%%%%%%%%%%%%%%%%%%%%%%%%%%%%%%
%%%%%%%%%%%%%%%%%%%%%%%%%%%%%%%

\section*{Geometry and surrogate model}
\begin{figure}[H]
\centering\includegraphics[width=0.8\textwidth]{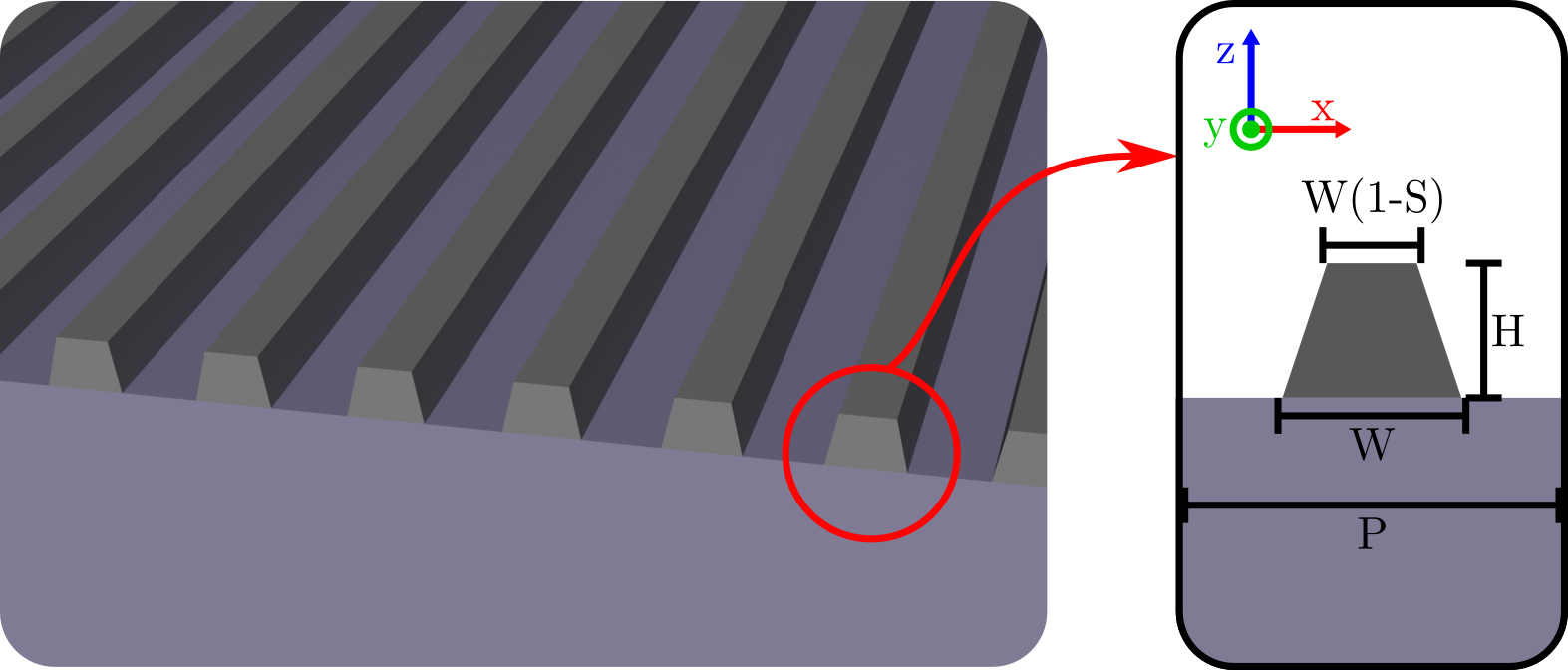}
\caption{Schematic representation of the simulated structure. The light is injected from top with normal incidence with electric field polarized along the x-direction. The inset refers to the geometrical parameters associated to the single unit-cell surrounded with periodic boundary conditions. } % Agreed with the changes
\label{fig:schem}
\end{figure}

%%%%%%%%%%%%%%5
%\subsection{Geometry}
%%%%%%%%%%%%%%%%%%%%%
Figure \ref{fig:schem} represents the considered metasurface geometry composed of slanted ridges of titanium dioxide (\ch{TiO2}) on top of a silicon dioxide (\ch{SiO2})  substrate with refractive index $\mathrm{n_s = 1.45}$. The refractive index of \ch{TiO2} is determined through ellipsometry (\textcolor{black}{available in the supplementary information section}). The inclusion of slanted gratings in the metasurface design introduces additional degrees of freedom, enabling the appearance of sharp resonances in the reflection spectrum. This is achieved by breaking the symmetry in the z-direction, resulting in high-quality resonance modes \cite{barton2021wavefront, lawrence2020high}. Four parameters, namely the metasurface period (P), resonator height (H), base width (W), and the ratio between the top and base widths of the resonator (S), are optimized in this study. % 
%%%%%%%%%%%%%%%%%%%%%%%%%%%%%%%%%%%%
%%%%%%%%%%%%%%%%%%%%%%%%%%%

The first step to train the \gls{ANN} is to generate a dataset. One important aspect that we considered to build it is the fabrication constraint that imposes a maximum aspect ratio of 2 between the height and width of the resonators. The dataset consists of \textcolor{black}{585} simulations  generated by uniformly sweeping the four parameters indicated in Figure \ref{fig:schem}. The period P ranges from 250 nm and 510 nm. For the height H, the range spans from 100 nm to 425 nm. In order to ensure that the ratios H/W and H/(P-W) remain less than 2, the width W varies between H/2 and \textcolor{black}{(2P-H)/2}, and therefore, the height H must obey the inequality H/2 < \textcolor{black}{(2P-H)/2}. Additionally, the parameter S ranges from 0 to 0.8, with intervals of 0.1.  The simulations were performed on an Intel® Xeon® W-2125 Processor operating at 4.0 GHz and allocating \textcolor{black}{4} threads, taking a total of \textcolor{black}{33 minutes and 40 seconds}.  The description of the 2D Python MEEP \gls{FDTD} simulations are described in the first section in the supplementary information.%

%%%%%%%%%%%%%%%%%%%%%%%%%%%%%%%%%%%%%%%%%%%%%%%%%
% \subsection{Surrogate Model}

Subsequently, a feedforward \gls{ANN} surrogate model was trained to forecast the reflection spectral response, considering the resonator's geometry as input. In this study, we leverage the surrogate model for two distinct purposes. Firstly, it enables rapid estimation of the reflection spectrum, achieving significant computational speed-ups compared to full wave \gls{FDTD} simulations. Secondly, the surrogate model is crucial in the inverse design process, playing a critical role in computing optimal solutions. As a result, the surrogate model's performance is of paramount importance, as it must deliver exceptional precision to provide the most accurate approximations possible. By fulfilling these requirements, the surrogate model accuracy significantly contributes to the effectiveness and success of this study. Our surrogate model is a classical \gls{MLP} model with fully connected layers. In this scenario, there are several ways to configure this \gls{MLP}, and we will compare two different strategies to map the input geometries into a spectrum.

\begin{figure}[H]
\centering\includegraphics[width=0.85\textwidth]{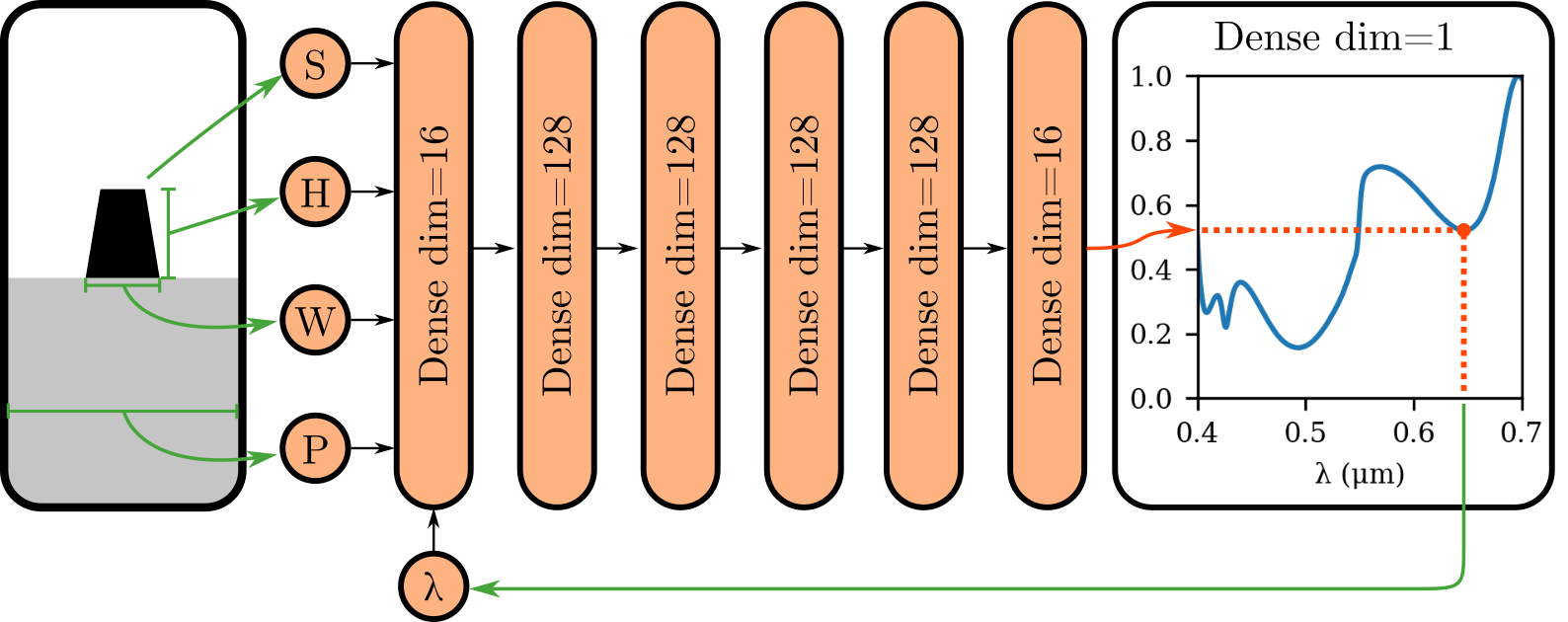}
\caption{The diagram represents the architecture of the surrogate model \gls{ANN}. The green arrows represent the input parameters. The surrogate model is designed to calculate a single output dimension, which corresponds to the reflection at the desired wavelength as shown by the red arrow. On the left side, a typical example of the resonator's geometry is provided, while on the right side, an example spectrum is displayed. It is important to note that these examples are purely for illustrative purposes and do not represent any real simulations. All the layers in this architecture are initialized using Glorot normal initializers as discussed in Ref. \cite{glorot2010understanding}. The model was trained using the TensorFlow framework, employing a holdout methodology \cite{abadi2016tensorflow}. The dataset is divided into two subsets: \textcolor{black}{555} simulations to update the weights of the \gls{ANN} and \textcolor{black}{30} simulations for validation. The optimizer of choice was Adam with \gls{MSE} loss and batch size of 512 \cite{kingma2014adam}. The training stopped at epoch \textcolor{black}{82} due to the lack of improvement of the validation loss within the last \textcolor{black}{20} epochs. The final \gls{MSE} loss was \textcolor{black}{$2.6996\times 10^{-4}$ and validation loss, $3.2745\times 10^{-4}$. The graph showing the loss evolution at each epoch can be found in section 2 of the supplementary information document}. The total training time was \textcolor{black}{9 min 48s on a Google colab's \gls{CPU}}.}
\label{fig:surrogate_schem}
\end{figure}

The first approach uses the geometric parameters as inputs and generates a multi-dimensional vector as output. Each dimension of this vector corresponds to a distinct wavelength in the reflection spectrum. The second approach involves incorporating the wavelength as an input parameter while outputting a single dimension representing the reflection amplitude at that particular wavelength (Figure \ref{fig:surrogate_schem}). Although both methods initially seems to be identical, the second is more convenient for this study. By incorporating the wavelength as an input, each point in the spectrum becomes a unique data point for the surrogate model. In contrast, the first method contains only \textcolor{black}{585} data points available for training the \gls{ANN}. Conversely, the second method significantly enhances the availability of data samples, yielding a total of \textcolor{black}{292,500} individual data points resulting from sampling the spectrum across 500 different wavelengths.

Figure \ref{fig:surrogate_schem} represents the architecture of the surrogate model. In general, we consider the \gls{GELU} activation function which is a gating activation with a continuous derivative. Considering \gls{GELU} implies the continuity of the output and its derivative, providing smoother response compared to the classical \gls{ReLU} activation \cite{hendrycks2016gaussian}. It is worth mentioning that the last layer contains a single linear neuron  to compute a singular output dimension aligning  with the reflection observed at the designated wavelength as indicated by the red arrow in Figure \ref{fig:surrogate_schem}. 

%%%%%%%%%%%%%%%%%%%%%%%%%%%%%%%%%%%%%%%%%%%%%%%%%%%
%%%%%%%%%%%%%%%%%%%%%%%%%%%%%%%%%%%%%%%%%%%%%%%%
 
\begin{figure}[H]
\centering\includegraphics[width=0.80\textwidth]{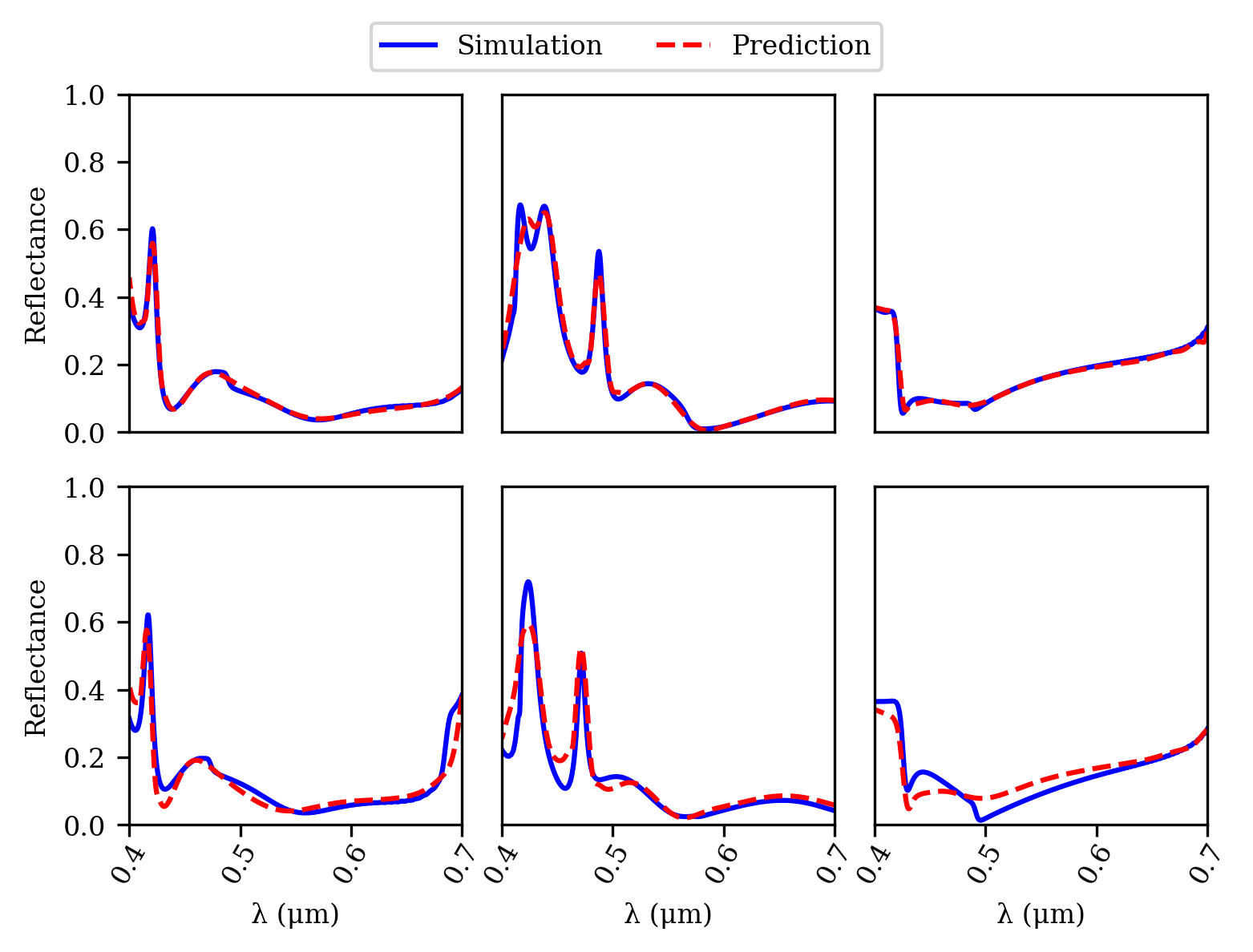}%{surrogate_examples.png}
\caption{Comparison between fullwave simulation and the prediction from the surrogate model. Top figures indicate the reflection spectra of three randomly selected samples from the dataset. The bottom figures corresponding to the same top geometries while introducing a small random deviation to the geometrical parameters. Notably, the surrogate model demonstrates robust generalization capabilities despite the inherent sparsity of the dataset.}
\label{fig:4layer}
\end{figure}

Figure \ref{fig:4layer} illustrates the generalization capacity of the surrogate model. Despite being trained on a relatively small dataset, the model is able to reproduce the spectra associated to the geometries that lie beyond the training dataset. This accomplishment can be attributed to the integration of wavelength as an input variable and the meticulous training methodology employed to mitigate overfitting. The demonstrated efficiency of this approach underscores the potential for constructing dependable surrogate models even when confronted with restricted training data.
%%%%%%%%%%%%%%%%%%%%%
%%%%%%%%%%%%%%%%%%%%%%%%

%%%%%%%%%%%%%%%%%%%%%%%%%%%%%%%%%% 
\section*{Inverse Design}
\begin{figure}[H]
\centering\includegraphics[width=0.8\textwidth]{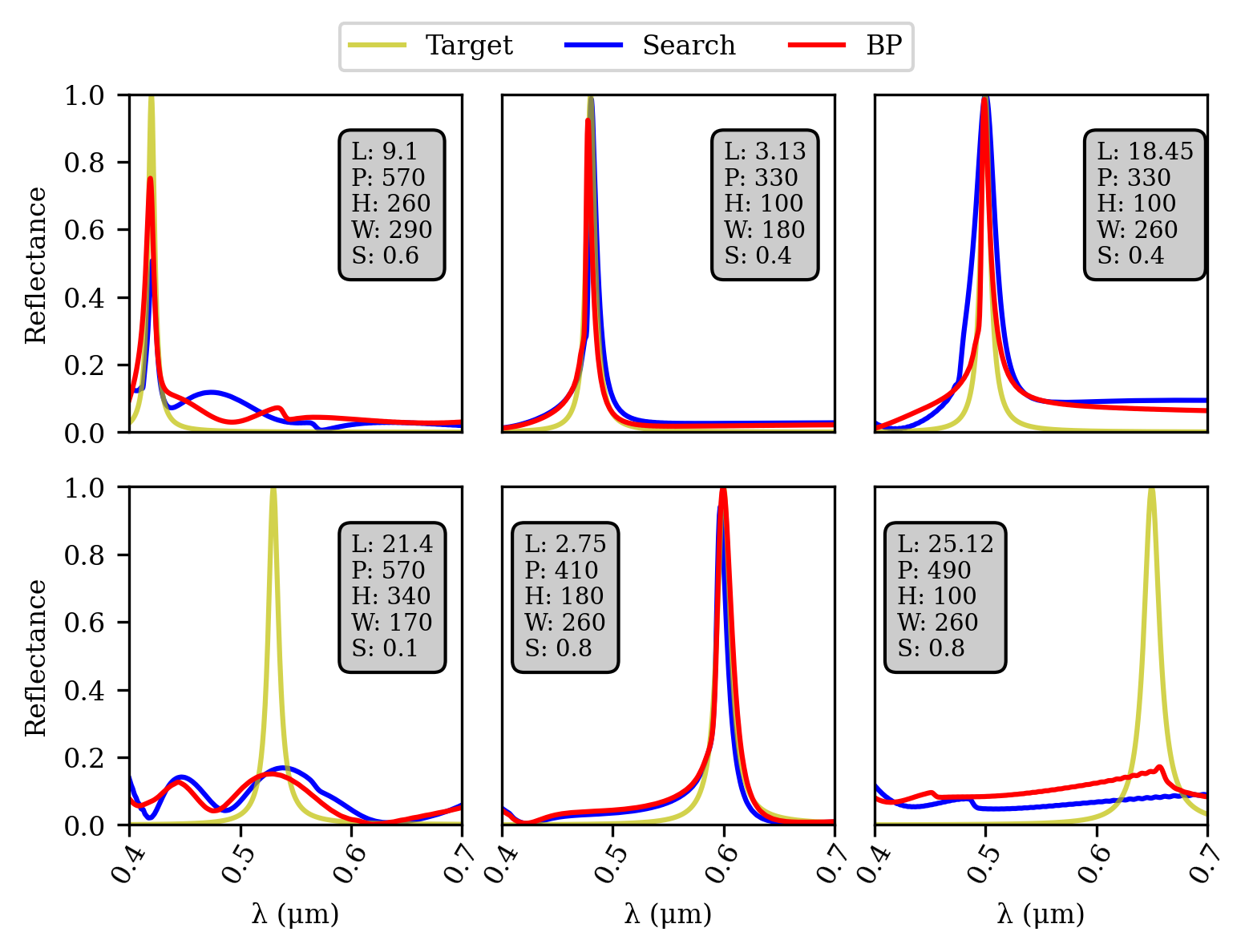}
\caption{Results from the dataset search using a Lorentzian lineshape in frequency as target for the dataset search for the wavelengths of \SI{420}{nm} \SI{480}{nm}, \SI{500}{nm}, \SI{530}{nm}, \SI{650}{nm} \SI{650}{nm} with $\omega=0.01$ from left to right, top to bottom. The metric used to evaluate the results was the \gls{MSE} (L) \textcolor{black}{in $10^{-3}$} between the target spectrum and the closest one in the dataset, while the geometric parameters (P, H, W and S) \textcolor{black}{of the device found on the search} were described in Figure \ref{fig:schem}. Blue and red curves represent the best design without and with back-propagation optimization on best design found in the dataset. A comparison of \textcolor{black}{two different resonance shapes with different  $\omega$ values are also tested in figure S5 provided in the supplementary information}.}
\label{fig:scan_comparison}
\end{figure}

In this particular section, an inverse design methodology is formulated for metasurface color filter. The primary objective is to determine the corresponding set of parameters based on the desired spectrum response. To initiate the process, the target line shape spectrum must be established, and in our research, a Lorentzian function shape resembling a Fano resonance-like shape is employed:
%%%%%%%%%%%%%%%%
\begin{equation}
\centering
        L\left(f\right) = \frac{\omega^2}{\omega^2 + 4\left(f - f_0\right)^2},
    \label{eq:lorentzian}
\end{equation}
%%%%%%%%%%%%%%%%%%%%%%%%%%%
where $\omega$ represents the full width at half maximum (FWHM) while $f_0$ corresponds to the central frequency. The units employed are in accordance with MEEP's configuration, where both frequency and FWHM are measured in $\mathrm{\mu m^{-1}}$. Additionally, it is worth noting that throughout the optimization process, the FWHM remains unchanged for all targets. This means that the Quality Factor (QF) of the desired spectra undergoes variations as a function of $f_0$, in accordance with Equation \ref{eq:quality}.
%%%%%%%%%%%%%%%%%%%%%%%%%%%
\begin{equation}
        QF := \frac{f_0}{\omega}
    \label{eq:quality}
\end{equation}
%%%%%%%%%%%%%%%%%%%%
\begin{figure}[H]
\centering\includegraphics[width=0.85\textwidth]{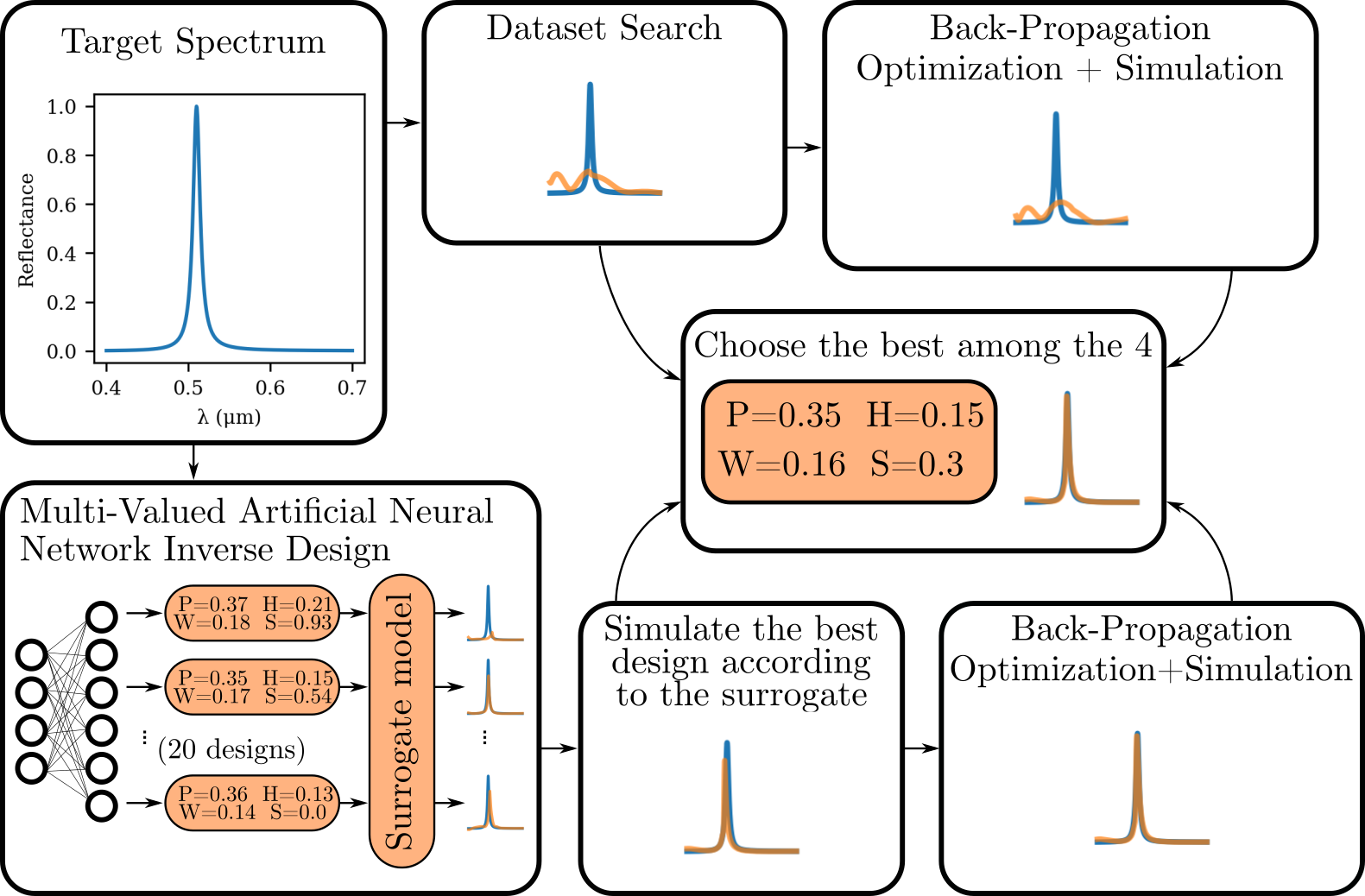}
\caption{Optimization diagram for the inverse design methodology, where the target reflection spectrum is specified in the cyan curve. To achieve the desired outcome, two complementary paths are identified. The first path involves exploiting the dataset scan and searching for the best design that will be subjected to a back-propagation optimization to fine-tune the parameters, as indicated in the top blocks. On the other hand, the second path involves passing the target through the \gls{MVNN} to obtain 20 designs. These designs are then validated using the surrogate model, and the solution that presented the least \gls{MSE} compared to the target is chosen. Finally, a back-propagation optimization is performed on that design. Ultimately, the best solution among the four selected designs from the two paths is chosen. Further details can be found in the text.}
\label{fig:opt_diagram}
\end{figure}
%%%%%%%%%%%%%%%%%%%%%%%%%
The optimization scenario illustrated in Figure \ref{fig:opt_diagram} depicts the inverse design process. The desired spectrum is represented by the cyan curve in the top left corner. In this stage, we employ two complementary approaches for inverse design. The first approach involves a straightforward search through a dataset, represented by the arrow pointing to the right of the target spectrum. For each simulation in the dataset, we calculate the \gls{MSE} loss between the target spectrum and the simulated spectrum. By doing so, we can identify the design that produces the lowest \gls{MSE}. Although this design is the best within the dataset, it fails to accurately match the desired response, as indicated in Figure \ref{fig:scan_comparison}. To address this limitation, we utilize a fully differentiable surrogate model and perform gradient optimization to finely adjust the parameters. As it can be seen in Figure \ref{fig:scan_comparison} running the back-prop\textcolor{black}{a}gation yields interesting designs. However, still this approach is not able to produce vivid colors along all the visible regime. Further the resonances are mainly associated with higher order modes appearing at shorter wavelengths (see last column) that allows for non-pure color. The optimized solution, along with the best design found within the dataset, are both stored for later comparison. \textcolor{black}{A detailed explanation of the methodology used to perform back-propagation is displayed in section 3 of the supplementary information.}
%%%%%%%%%%%%%%%%%%%%%%%%%%%%%%%%%%%%%%%%%%%%%%%%%%%%%%%%%%%
%%%%%%%%%%%%%%%%%%%%%%%%%%%%%%%%%%%%%%%%%%%%%%%%%%%%%%%%%%%%%%%%%%

The second complementary approach, indicated by the downward arrow from the target spectrum in Figure \ref{fig:opt_diagram}, involves the utilization of a \gls{MVNN} to generate robust solutions for the inverse design problem. A \gls{MVNN} is essentially an \gls{ANN} capable of producing multiple solutions based on a single input. In the context of our study, the adoption of the \gls{MVNN} proves highly advantageous as it effectively addresses the challenge of multiple parameter responses corresponding to a given objective target \cite{zhang2018multivalued}. To ensure reasonable accuracy, we choose to employ a model that yields 20 outputs. The detailed architecture and training procedure of the \gls{MVNN} are provided in the supplementary information section 5.
After acquiring the set of 20 distinct designs, a validation process is conducted using the surrogate model. Among these designs, the one that exhibits the lowest \gls{MSE} loss when compared to the target is initially stored for future comparison. To further improve the results, this best design is subjected to back-propagation optimization. By employing a combination of the dataset search method indicated by the top blocks in Figure \ref{fig:opt_diagram} and the \gls{MVNN} approach, followed by back-propagation indicated by the bottom blocks in Figure \ref{fig:opt_diagram}, we successfully identify four robust designs. This entire process, encompassing both approaches, takes approximately 100 seconds to complete. Finally, the design exhibiting the lowest \gls{MSE} loss compared to the target is selected as the final choice. \textcolor{black}{We refer to section 6 in the supplementary information for more details regarding the performance of \gls{MVNN} as a function of the dataset size. Further, a detailed comparison between \gls{MVNN} and a state-of-the art algorithm can be found in section 7 in the supplementary information.}

\begin{figure}[H]
\centering\includegraphics[width=\textwidth]{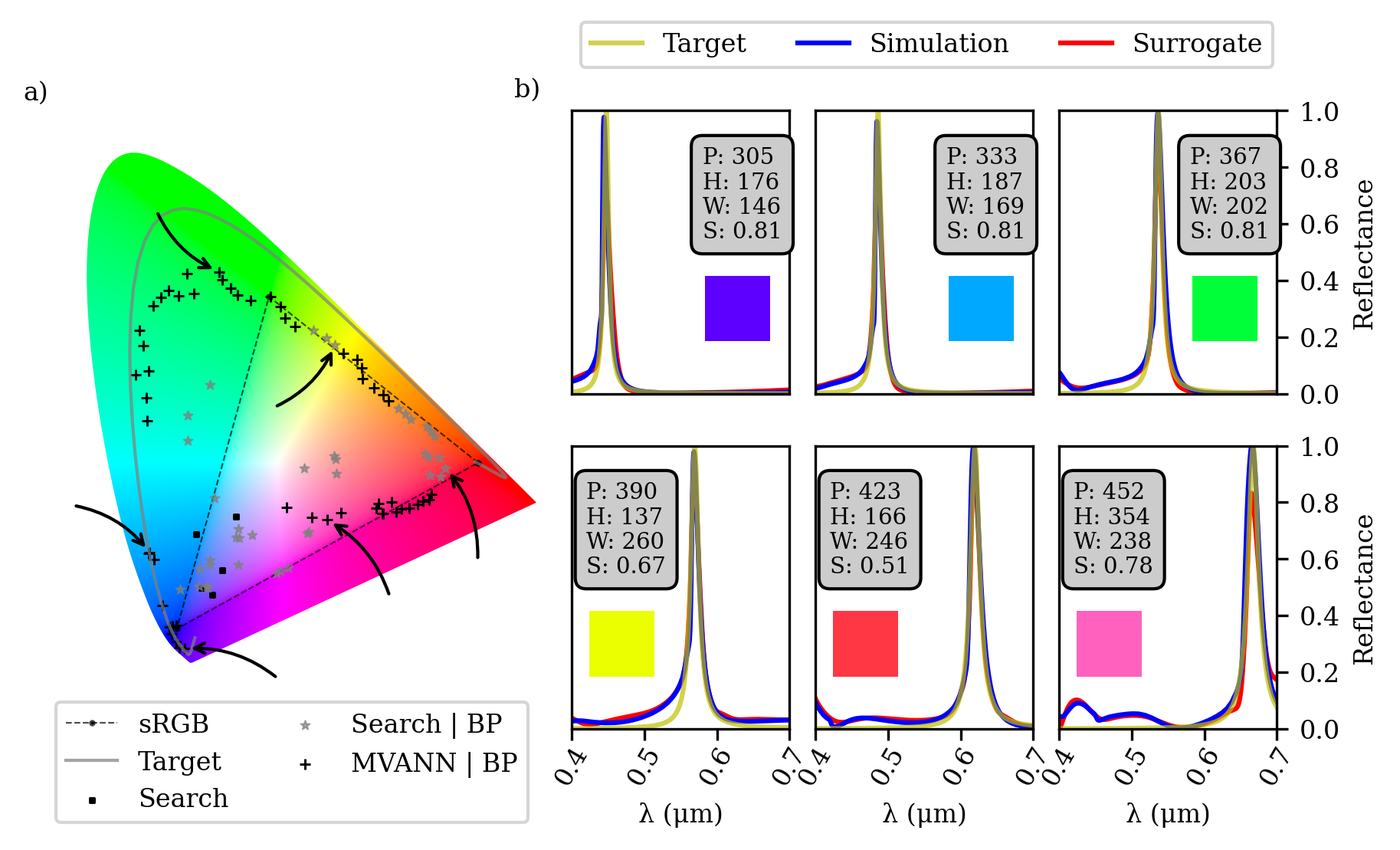}
\caption{(a) Chromaticity diagram depicting the outcomes of the 100 optimization results. The arrows in the diagram indicate the corresponding designs showcased in (b). The gray line represents the calculated perceived color corresponding to the target spectrum. (b) detailed examination of 6 different optimizations, displaying their dimensions (in nm) of each geometrical parameter and the corresponding sRGB color representation of the simulated spectrum. }
\label{fig:color_opt}
\end{figure}

%%%%%%%%%%%%%%%%%%%%%%%%%%%%%%%%%%%%%%%%%%%%%%%%%%%%%%%%%%%%%%%%%%%%%%%%%%
By employing the aforementioned inverse design methodology illustrated in Figure \ref{fig:opt_diagram}, we perform a total of 100 optimization iterations, as depicted in Figure \ref{fig:color_opt} (a). These iterations involve sweeping the desired wavelength across the range of \SI{400}{nm} to \SI{700}{nm}, thereby encompassing the entire spectrum of colors. The target function is defined as a Lorentzian spectrum with a Full Width at Half Maximum (FWHM) value of $\omega = 0.01$. Subsequently, we calculate the perceived color values (x and y) on the chromaticity diagram using the CIE 1931 $\mathrm{2^\degree}$ standard observer color matching functions and the "Equal energy" illuminant, which ensures that incident light of all wavelengths possesses uniform power \cite{schanda2007colorimetry}. As depicted Figure \ref{fig:color_opt} (a), we successful optimized metasurface designs that generate colors closely resembling our target (gray curve). Consequently, a significant number of these optimized designs fall outside the sRGB zone, represented by the dashed triangle in Figure \ref{fig:color_opt} (a), expanding the color gamut achievable with metasurfaces beyond the limitations of conventional display technologies.

 In Figure \ref{fig:color_opt} (b), we select  six distinct designs to demonstrate the spectral response of our optimized configurations, as indicated by the black arrows in Figure \ref{fig:color_opt} (a). Our inverse design approach has led to metasurface designs that exhibit a reflection spectrum response that is consistent with the target Lorentzian spectrum across all desired wavelengths. Moreover, the background response in the spectrum is nearly flat, providing a vibrant color response for our optimized configurations. \textcolor{black}{A typical example of the field profile is presented in the supplementary information}. Additionally, a numerical comparison between FDTD calculations and COMSOL has been included to verify our results. Furthermore, we have conducted a full-wave simulation with finite structure to identify the number of required periods to retrive the results of a single unitcell simulation with periodic boundary conditions (for more details, the reader can refer to the supplementary information). To the best of our knowledge, the optimized metasurface designs yield the most vivid color filter reported so far in the literature \cite{liu2022transfer, panda2020transmission, wang2022tunable, khaidarov2022large, hemmatyar2019full}. It is worth mentioning that the entire optimization process took \textcolor{black}{2 hours, 37 minutes, and 45 seconds. Among this time, 1 hour, 26 minutes, and 20 seconds were dedicated to simulations, 1 hour, 2 minutes, and 6 seconds were allocated for back-propagation optimization, 8 minutes and 27 seconds were spent on predictions using both the surrogate model and the \gls{MVNN}, and 250 milliseconds were utilized for searching the best design within the dataset.}
% %%%%%%%%%%%%%%%%%%
%%%%%%%%%%%%%%%%%%%%%%%%%%%%%
%%% height fixing

The results presented in Figure \ref{fig:color_opt} show that the optimized designs have varying heights. However, considering the complexity associated with fabricating structures of different heights, it is preferable to find designs with a fixed height. Therefore, we introduce a constraint to the back-propagation optimization process, requiring all designs to have equal heights. To achieve this, we modified the inverse design workflow as follows: Initially, we obtained the lowest 5 losses from the \gls{MVNN} for each of the 6 different target designs, each representing a different color and having a different height. Subsequently, we identified designs that were closer in their height values. We calculated the mean height of these selected designs for each color and used this mean value as an initial guess for optimizing each target using the back-propagation optimization process.

To accommodate the height constraint, we introduced a dedicated layer in the surrogate model. This layer connected the height input solely to the initial guess and consisted of a single weight and bias. This shared layer was employed throughout all optimization iterations. By optimizing all 6 models simultaneously within the same batch, we ensured that the loss for each individual model progressively decreased with each training epoch. Consequently, we transformed the fixed height constraint into an optimized parameter, allowing us to determine the most suitable height value for achieving the desired outcomes across all six colors.

%%%%%%%%%%%%%%%%%%%%%%%%%%%%%%%%%%%%%%%%%%%%%%%%%
\begin{figure}[H]
\centering\includegraphics[width=\textwidth]{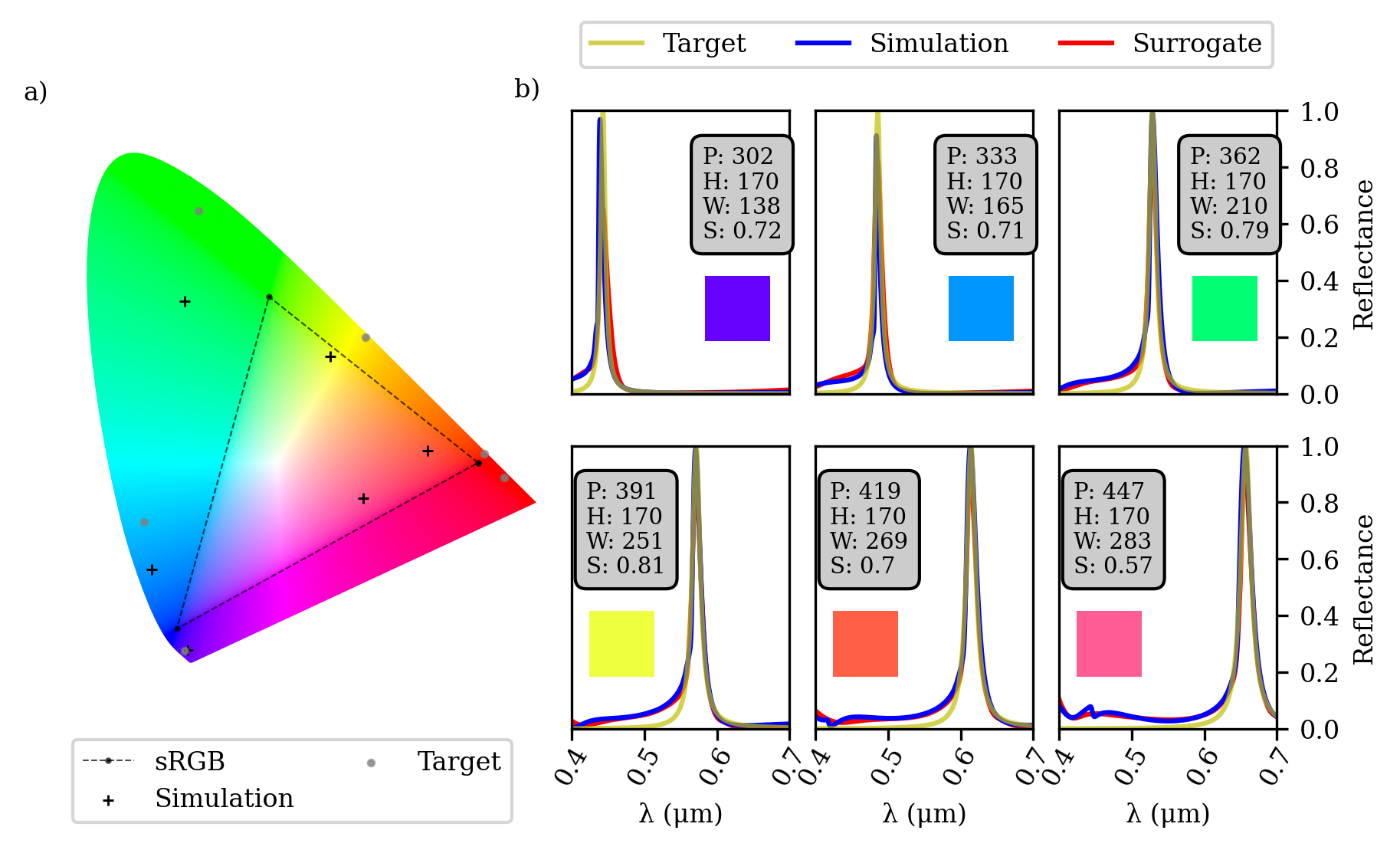}
\caption{Optimization results based on a fixed height configurations. (a) refers to the chromaticity diagram of 6 optimized designs.  The simulated spectra is given in (b) \textcolor{black}{and the time used for training, and  optimizing is depicted in the supplementary information section.}}
\label{fig:opt_6_fix_h}
\end{figure}
%%%%%%%%%%%%%%%%%%%%%%%%%%%%%%%%%%%%%%%

\begin{figure}[H]
\centering\includegraphics[width=\textwidth]{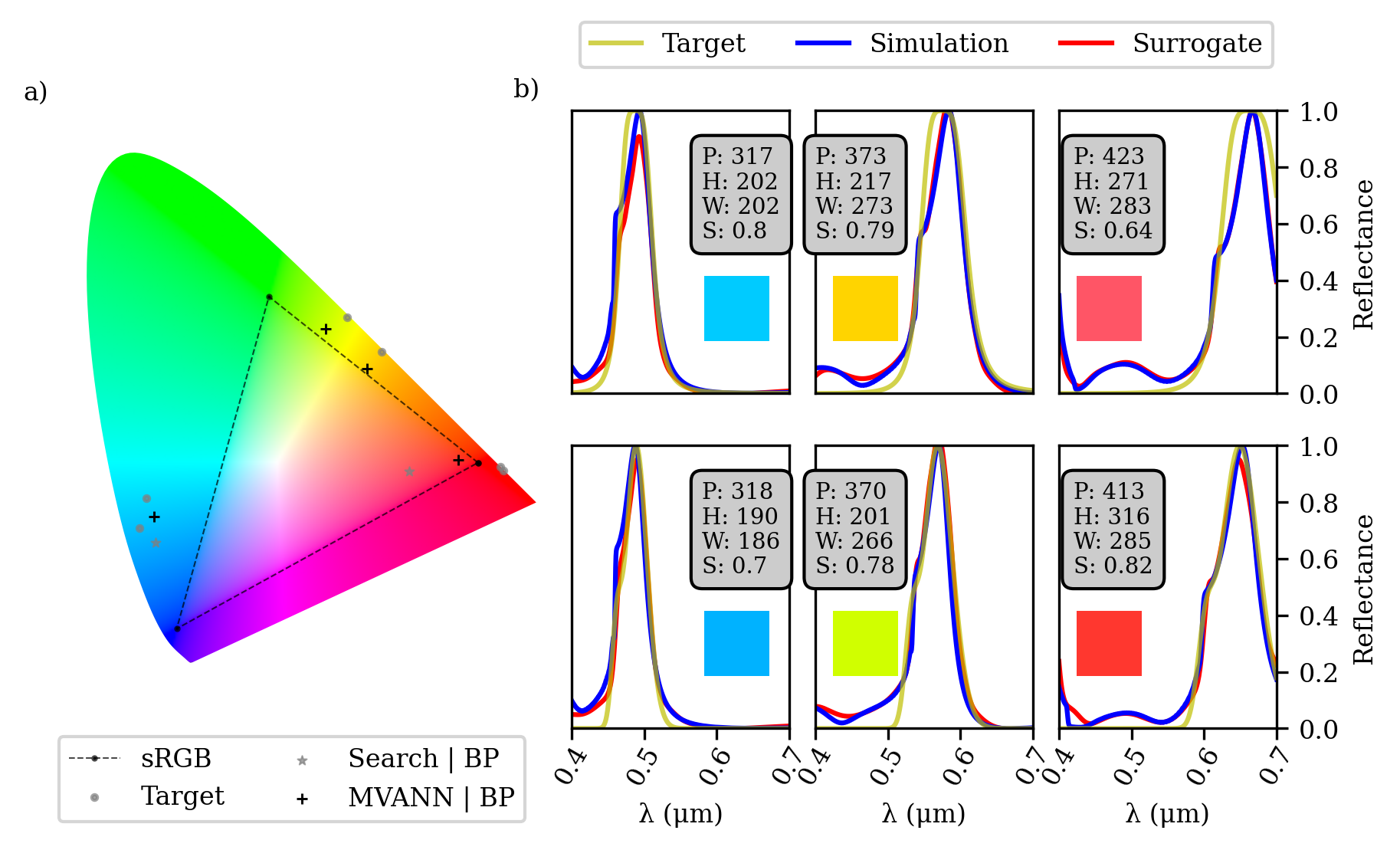}
\caption{Optimization results targeting a broader Lorentzian ($\omega^2/(\omega^2 + 4\left(f - f_0\right)^6)$) on the top row and a combination of Gaussians to mimic a strong secondary resonance on the bottom row. (a) refers to the chromaticity diagram of 6 optimized designs. The simulated spectra is given in (b). It can be seen that we achieved the most vivid red.}
\label{fig:585_shapes_opt}
\end{figure}

The results presented in Figure \ref{fig:opt_6_fix_h} demonstrate the effectiveness of our approach in achieving favorable outcomes for most of the considered colors.
%%%%%%%%%%%%%%%%%%%%%%%%%%%%%%%%%%%%%
\textcolor{black}{Our optimization approach allows us to explore various line shapes using the same dataset, as depicted in Figure \ref{fig:585_shapes_opt}. Notably, our algorithm excels at identifying optimized designs that seamlessly match the desired response. It is important to mention that our study focuses on achieving pure colors across the entire spectrum, rather than emphasizing a single color. Yet, interestingly, our algorithm is capable of optimizing designs to replicate a vibrant red color response, a task that traditionally involves complex optimization processes (see \cite{lin2023efficient,dong2022schrodinger}). However, our design approach, coupled with advanced inverse design techniques, simplifies the process of identifying designs that mimic this vivid red color response, all while relying on the same dataset. Furthermore, it is worth noting that prior studies employed half the number of simulations to optimize a single color compared to our methodology.}
%%%%%%%%%%%%%%%%%%%%%%%%%%%%%
%%%%%%%%%%%%%%%%%%%%%%%%%%%%%%%

\section*{Conclusion}

In conclusion, our research presents a novel \gls{DL} methodology for optimizing resilient designs of vivid color filter metasurfaces. By utilizing a surrogate model constructed from a dataset of only \textcolor{black}{585} simulations, our approach demonstrates exceptional efficiency in the optimization process. The numerical tool developed in this study enables the cost-effective fabrication of structural color filters by exploring a wide range of narrow line shapes that exhibit high-quality resonances, aligning with desired spectral reflection responses. Notably, our methodology expands the color gamut beyond the conventional RGB colors, offering unprecedented versatility in color generation. Furthermore, our \gls{DL} approach successfully respects fabrication constraints, ensuring practical feasibility. The achievements of our research significantly contribute to the field of optical device design. By pushing the boundaries of metasurface optimization, we open up new possibilities for the development of advanced optical devices. The proposed methodology holds promise for various applications, such as display technologies, data encoding, and artistic expression. These notable advancements not only enhance the understanding of metasurface design principles but also provide valuable insights for future research endeavors.

%Bibliography
\bibliographystyle{unsrt}  
\bibliography{references} 

\section*{Author contributions statement}

 A.C.D developed the DL architecture and performed the numerical simulation. ME proposed the structure and coordinated the work. M.E, A.C.D, S.L discussed the numerical results. M.E, A.C.D, H.E.H.F, and M.A discussed the physical results. All authors contributed to the writing and reviewed the manuscript. The authors declare that they have no  competing  interests.

\section*{Acknowledgment}
Hugo EHF would like to thank the support of Brazilian government agencies CAPES and CNPq (through project: 312714/2019-0), and the São Paulo Research Foundation (FAPESP) agency through the projects: 2021/11380-5 (CCD/CPTEn) and 2021/06506-0.

\end{document}

\maketitle

\section{Simulation setup}

\begin{figure}[H]
    \centering
    \includegraphics[width=0.9\textwidth]{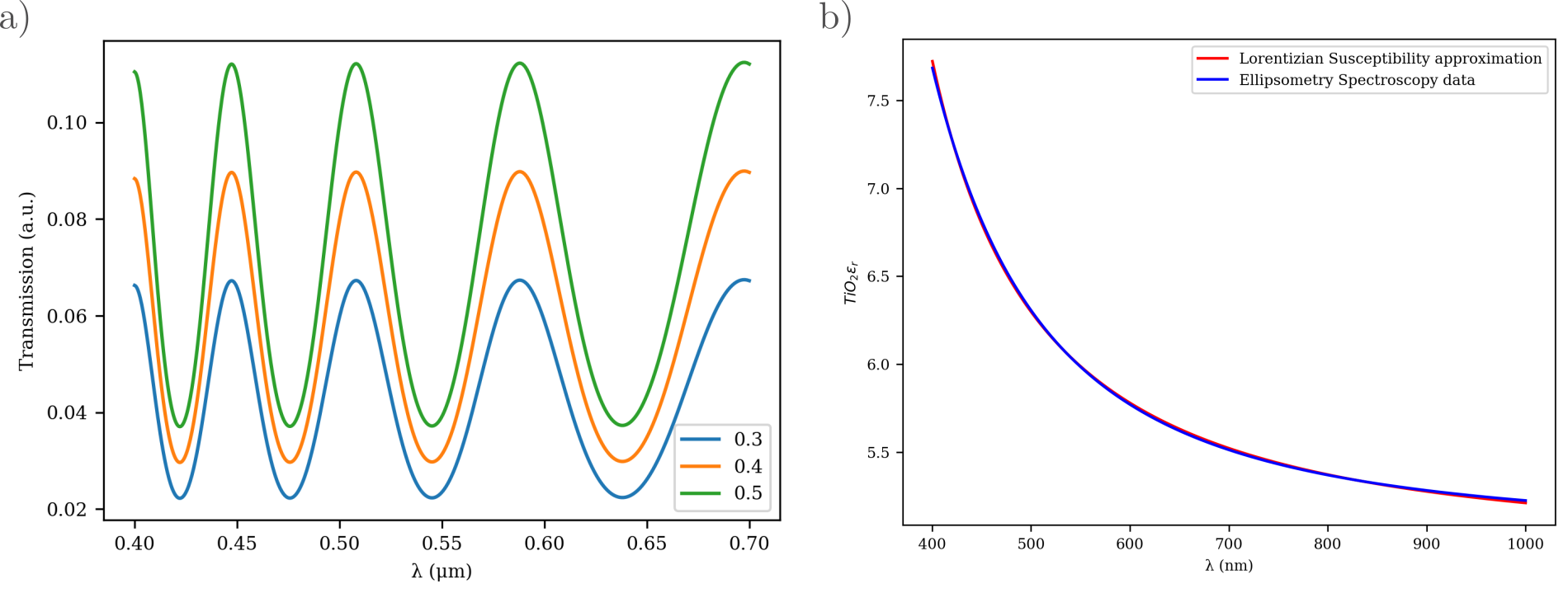}
    \caption{a) Transmission spectrum without any structure. Each curve represent a different resonator period. b) Comparative analysis of the measured dispersion of \ch{TiO2} and the approximation generated through modeling using MEEP. The experimental data capturing the dispersion characteristics were obtained through Ellipsometry spectroscopy detailed \textcolor{black}{below}. Subsequently, the collected data were fitted to a Lorentzian susceptibility function. Notably, the approximation achieved through the modeling process closely aligns with the measured dispersion, demonstrating the efficacy of the approach in accurately representing the dispersion behavior of \ch{TiO2}} % This is terrible I need to refine this image
    \label{fig:a}
\end{figure}

The simulations are realized using python's MEEP library \cite{oskooi2010meep}.The default resolution is 150 pixels per \SI{}{\um}, and therefore, 105 pixels per longest wavelength. However, if the cell size of the simulation is not an integer number of pixels, instead of adapting the resolution, the default MEEP procedure is to change the cell size. In ou\textcolor{black}{r} case, we require fine control over the period. That is why we instead used a value close to \SI{150}{nm} that resulted in the cell dimensions to be an integer number of pixels. MEEP stills warns that the grid volume is not an integer number of pixels and changes slightly the cell size due to the truncation of some values.
Each simulation only evaluates a single resonator and there is periodic boundary condition on each boundary of the simulation.
\gls{PML} layers with \SI{0.8}{\um} length were positioned on the extremes of the simulation except from the sides. MEEP is known to have issues regarding periodic boundary conditions and \glspl{PML}, and the longer the interface between them, the more artifacts the simulation acquires and easier it becomes to diverge. The other option was using Absorber layers, the problem is that they require much more space to fully extinguish the fields and increases the simulation time. Some reaching 20 minutes. As the length of the \gls{PML} is slightly larger than the longest wavelength, it \textcolor{black}{extinguishes} the fields while having imperceptible simulation artifacts. The same simulation that would last 20 minutes with the absorber layer, now takes only 14 seconds.
The total height of the simulation is \SI{3.1}{\um}. The height of the cell is adjusted if the resolution is updated.
The light is injected from top with normal incidence with electric field polarized in the x-direction.
There are 5 uniformly distributed in frequency Gaussian sources to guarantee that floating point errors does not interfere with the results.The transmission spectrum was sampled from \SI{400}{nm} to \SI{700}{nm} with \SI{500}{} points on the bottom of the simulation. An electric filed sampling on the x-direction was positioned between the sources and the middle of the simulation. It records the maximum value it ever encounters and then it stops the simulation if the field recorded on that point is $10^{-8}$ \textcolor{black}{smaller than} the maximum value it found. It then performs a \gls{FFT} to compute the transmission spectrum. As there are no losses in the simulation, the reflection spectrum is calculated using the transmission spectrum. In order to calculate the transmission spectrum, firstly it is required to validate the "raw" transmission \textcolor{black}{without any structure. Therefore, computing the spectrum of the source}. \textcolor{black}{By} varying the period of the resonator, \textcolor{black}{we observe} changes the intensity of the transmission as seen in \Cref{fig:a} a).
%%%%%%%%%%%%%%%%%%%%%
%%%%%%%%%%%%%%%%%%%%%%%%%%
\subsection*{\ch{TiO2} Ellipsometry Spectroscopy}

The characterization of the two-dimensional flat layers made of \ch{TiO2} was performed in order to
determine their optical constants. The measurements were carried out with a Wollam
M200V ellipsometer (350-1000 nm) on a layer obtained via sol-gel chemistry and dip-coating on a bulk Si substrate\cite{bindini2017}. 
The ellipsometric data ($\Delta$ and $\Psi$) are fitted by a Cauchy model to extract refractive index n and extinction coefficient k (being the latter equal to zero in the considered wavelength range).  
These layers are of optical quality being their surface roughness below 0.1 nm \cite{chehadi2021}. 

Measuring the refractive index evolution of a \ch{TiO2} thin layer kept in a ellipsometry-porosimetry chamber with controlled humidity (atmosphere where air and water were progressively mixed) we assess the adsorption/desorption of water. From the corresponding isotherms we obtain the mesoporosity of the layers\cite{loizillon2019} (pore
volume, size distribution, and interconnection through
adapted Kelvin models). Measurements on dense \ch{TiO2} (n $>$ 2.4) account for a porosity well below 10$\%$

The refractive index of the \ch{TiO2} was fitted by a Lorentzian susceptibility function using scipy's optimize curve\_fit function with initial guess 1.1 for epsilon, 3.85 of frequency and 3.75 of $\sigma$. The final parameters values are 1.87681265 for epsilon, 3.63269272 for frequency and 3.08109585 for $\sigma$. The difference between the ellipsometry data and the fitte\textcolor{black}{d} curve can be seen in \Cref{fig:a} b).

To calculate the spectrum source for every period, we assumed a linear correlation between the period and the magnitude of the transmittance, and performing a linear interpolation for every point in the spectrum, obtaining the source vector $\mathbf{a}$. The linear coefficient was close to 0 for all 500 points, and therefore, considered to be exactly 0. And therefore, to obtain the source spectrum of a resonator with period P, we can simply multiply the source vector $\mathbf{a}$ with the period of the metasurface.

\begin{figure}[H]
    \centering
    \includegraphics[width=\textwidth]{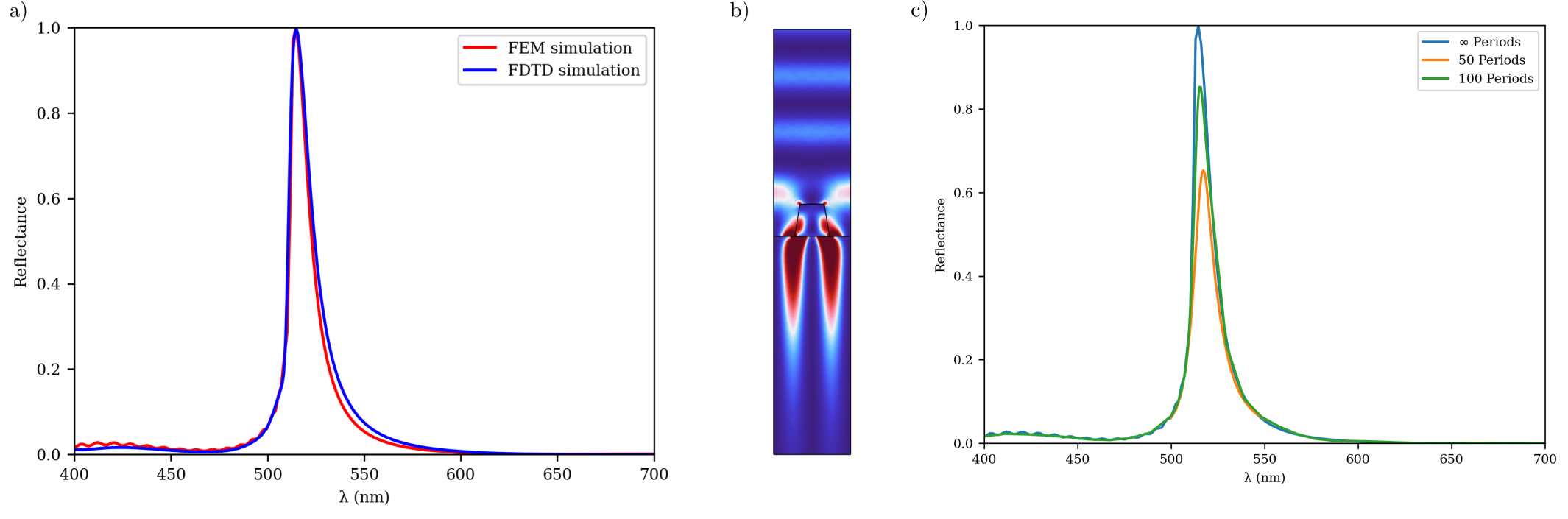}
    \caption{Comparison between the \gls{FDTD} MEEP simulations and \gls{FEM} COMSOL simulations. With P = \SI{353}{nm}, H = \SI{147}{nm}, W = \SI{159}{nm}, S = 0.3. a) Reflection spectrum and b) Electric field normalized at the wavelength corresponding to the peak reflactance. c) Difference between different array sizes.} % This is terrible I need to refine this image
    \label{fig:sup_info_comp}
\end{figure}

A comparison of MEEP \gls{FDTD} simulation and COMSOL multiphysics \gls{FEM}, has been and and shows on \Cref{fig:sup_info_comp} a) and b).
%%%%%%%%%%%%%%%%%%%%%

\begin{figure}[H]
   \centering
   \includegraphics[width=0.8\textwidth]{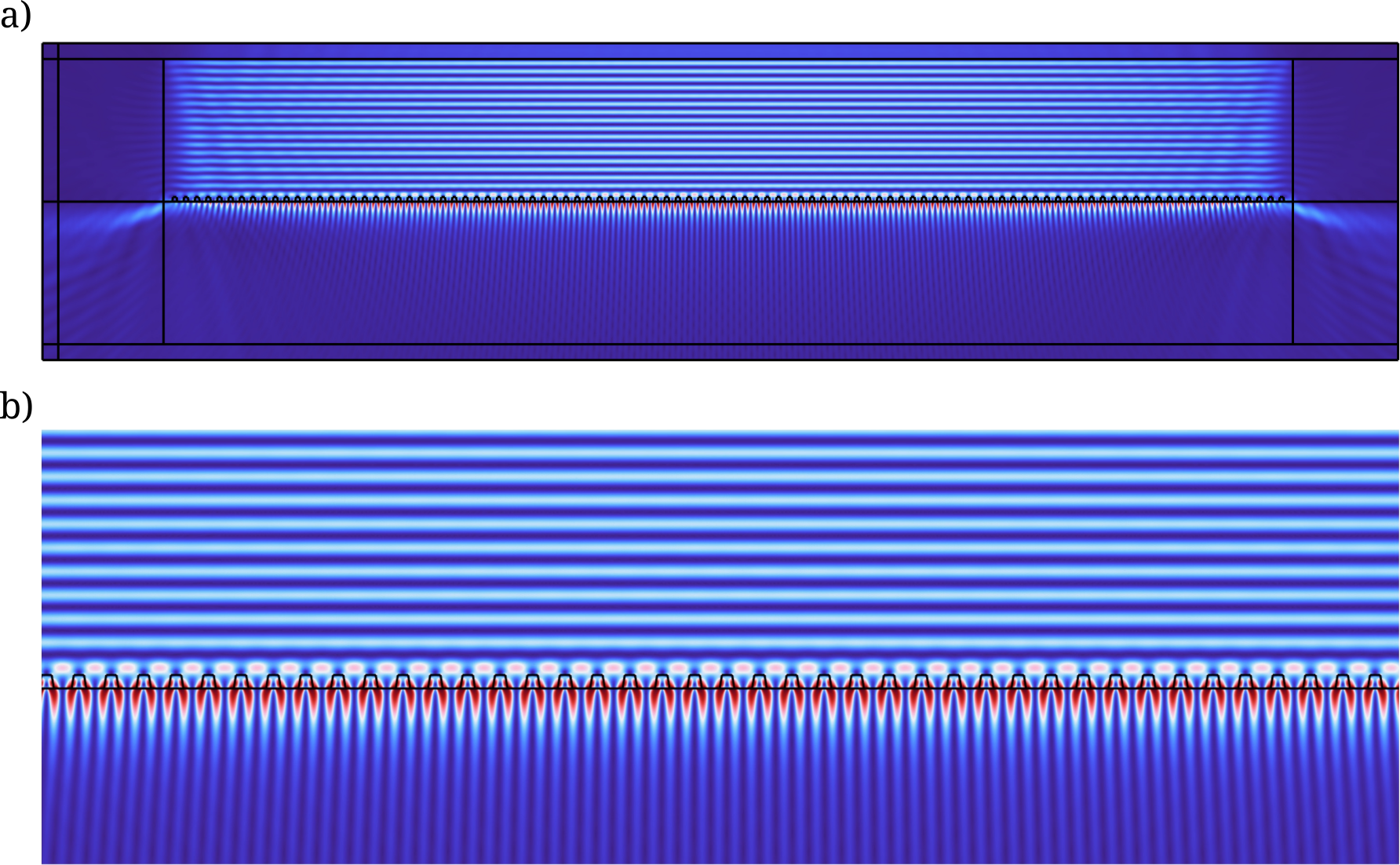} %fields_100.png
   \caption{Fields computed at the wavelength of \SI{515}{nm} for the scenario of 100 repetitions. a) Electric field normalized of the entire simulation domain. b) Zoom at the center of the simulation domain to obtain a better view of the local fields near the resonators.}
   \label{fig:fields}
\end{figure}

Figure \ref{fig:sup_info_comp} c) shows how the reflectance, transmittance and loss responds to different array sizes. As the period is \SI{353}{nm}, the last configuration with 100 repetitions is \SI{35.3}{\um} in length. And as seen, the resonance requires many repetitions to achieve the desired performance. The reason for this behavior is the mode excited by the resonant frequency. Which propagates in the x-axis in the substrate as seen by \Cref{fig:fields}.

\section{Surrogate Training}

\begin{figure}[H]
\centering\includegraphics[width=0.8\textwidth]{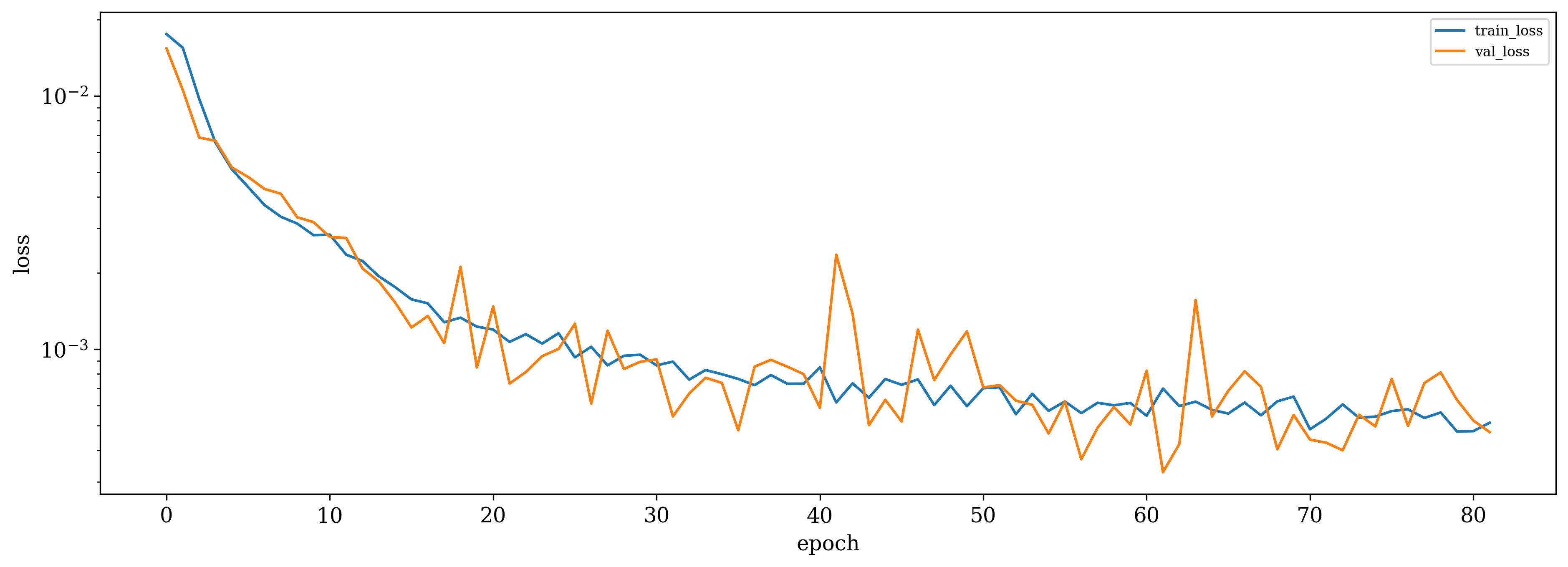}
\caption{\gls{MSE} loss of the training and validation sets in function of the epoch for the surrogate model.
}
\label{fig:surrogate_training}
\end{figure}

Figure \ref{fig:surrogate_training} displays the loss evolution at each epoch.

\section{Back-propagation optimization methodology}

By disabling the training of the weights of the surrogate model and added one layer before the input. The added layer has 5 inputs and the weights were initialized with the identity. The 4 first inputs (geometry parameters) are fully connected with the first 4 inputs of the surrogate model and the last input ($\lambda$) is solely connected to the surrogate model input referent to the wavelength. The training of this connection is disabled. The activation function for the described layer was \gls{ReLU} to prevent negative values. The most important part is the training. We used full batch Adam optimizer with initial learning rate of $10^{-5}$ with exponential decay with 1000 decay steps and decay rate of $0.98$. These choices proved necessary due to the needed small adjustments and the gradients grow after passing through the entire \gls{ANN} without any \gls{BN} layer. The training spanned 1800 epochs with patience of 6 with minimum delta of $10^{-6}$, meaning if the loss decrease less than $10^{-6}$ in 6 epochs, the training stops.
%%%%%%%%%%%%%%%%%%%%%
\textcolor{black}{We present the Python code using tensorflow and keras framework to run the back-propagation optimization.}
\begin{python}
lr_schedule = tf.keras.optimizers.schedules.ExponentialDecay(
                            initial_learning_rate=1e-5,
                          decay_steps=1000,
                          decay_rate=0.98)  # Learning rate schedule
opt = tf.keras.optimizers.Adam(learning_rate=lr_schedule)   # Using Adam

# Patience
callback = tf.keras.callbacks.EarlyStopping(monitor='loss', 
                                            patience=6, 
                                            min_delta=1e-6) 

# Define the model to be trained
Input_all = Input(shape=(5,))   # Input that will receive the initial guess

# Layer responsible for finding the geometrical parameters.
d1 = Dense(4, kernel_initializer='identity', activation='relu')(Input_all[:,:-1])

# Layer that will pass the wavelength.
# Note that we do not want to optimize this layer.
d2 = Dense(1, kernel_initializer='identity', trainable=False)(Input_all[:,-1:])     
d3 = Concatenate()([d1,d2])
inverse_designer = Model(inputs=Input_all, outputs=d3)

Surrogate_model.trainable = False   # Freeze the weight of the surrogate model.
Full_model_outputs = Surrogate_model(inverse_designer.output)

# Creating the model comprising the surrogate and our layer
Full_model = Model(inverse_designer.input, Full_model_outputs)  
Full_model.compile(optimizer=opt, loss='mse')

# Initial guess here can be from the MVANN or from the Search
# Its simply a list in the following format: [P, H, W, S]
ipt_i_tolist = initial_guess.tolist()   

# We then create a dataset comprising the combination of 
# same input and all the wavelengths
# wl is a list of all wavelengths
eng_inp = np.array([ipt_i_tolist + [w] for w in wl])    
                                                       
# The next step is training the new layer. Note that we want full batch optimization
# t_f is the target response
history = Full_model.fit(eng_inp, 
                        np.array(t_f), 
                        epochs=1800, 
                        callbacks=[callback],
                        verbose=0, 
                        batch_size=500)

# Then, to obtain the optimized result, we call it.
bp_result = inverse_designer.predict(eng_inp, verbose=0)     
\end{python}
%%%%%%%%%%%%%%%%%
\textcolor{black}{We decided to incorporate an additional layer for optimization rather than directly computing and applying gradients to the design space. This decision was primarily driven by the fact that the framework already had features like early stopping and learning rate schedules in place, which we found beneficial for our approach.}
%%%%%%%%%%%%%%%%%%%%%%%%%%%%%

\section{Optimization for lower quality factor resonance targets}

\begin{figure}[H]
\centering\includegraphics[width=0.8\textwidth]{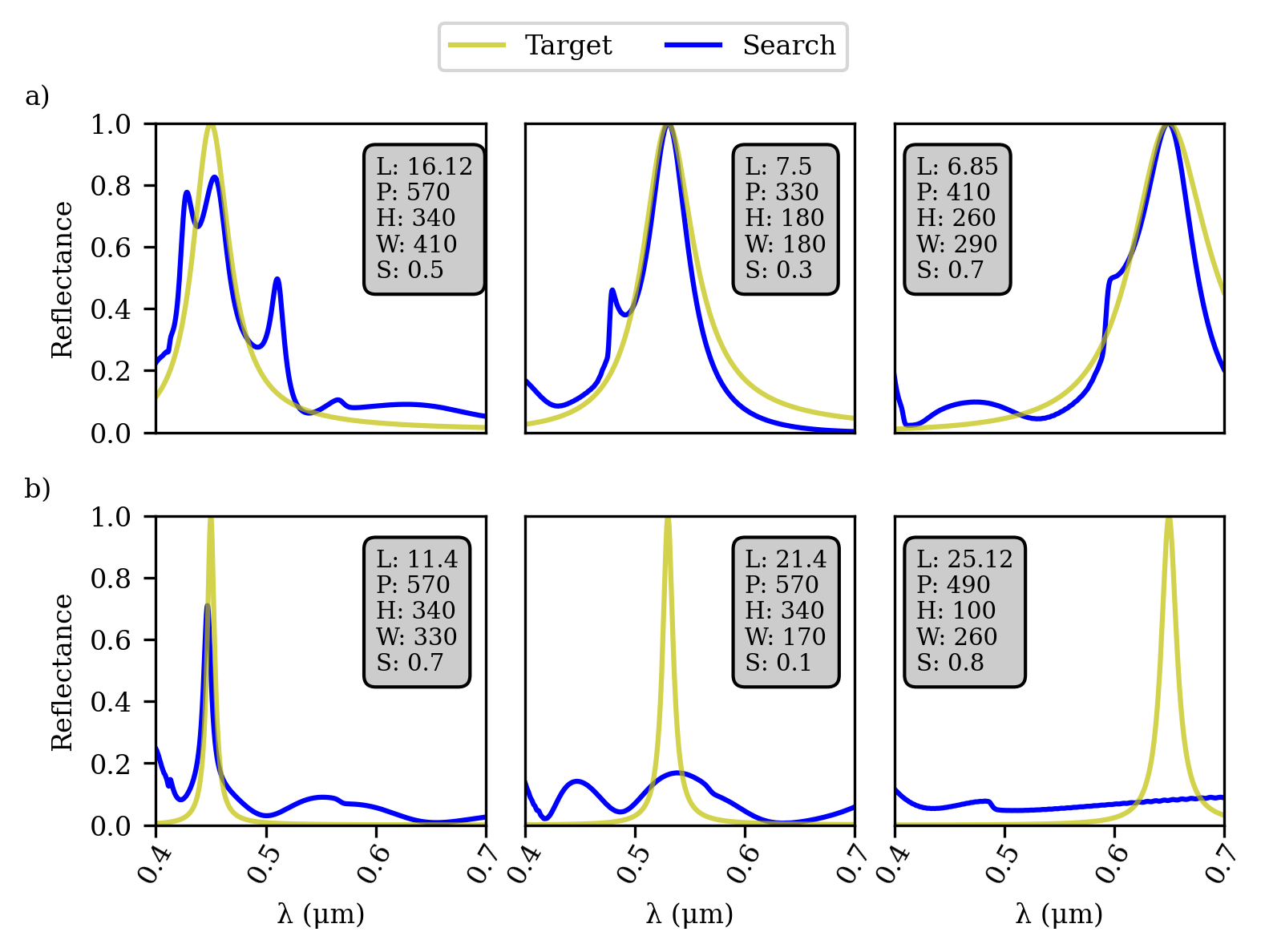}
\caption{Results from the dataset search using a Lorentzian lineshape in frequency as target for the dataset search for the wavelengths of \textcolor{black}{\SI{450}{nm}}, \SI{560}{nm} and \textcolor{black}{\SI{650}{nm}} with a) $\omega=0.05$ and b) $\omega=0.01$. L is the \gls{MSE} between the target spectrum and the closest one in the dataset, and P, H, W and S are the geometric parameters. As seen, unlike the low quality target objectives give in (a), the dataset search could not find a good solution when targeting sharp resonances as it is depicted in b).
}
\label{fig:scan_comparison}
\end{figure}

By performing the dataset search with relaxed objectives that concerns a broader \gls{FWHM}, a reasonable solution can be found on the dataset. Turning the Remaining optimization methodology unnecessary as seen by \Cref{fig:scan_comparison}.

As our objective is relaxed, there is no need to refine the solutions found by searching the dataset.

\section{MVANN}

A \gls{MVNN} is simply a \gls{ANN} capable of outputting multiple solutions given a single input. The adoption of the \gls{MVNN} in this context is particularly advantageous due to its ability to mitigate the issue of multiple parameter responses for a given objective target \cite{zhang2018multivalued}. This feature ensures that the \gls{MVNN} provides reliable and consistent results in non-unique scenarios where multiple solutions can arise from a single input target spectrum.

One of the greatest challenges even for a \gls{MVNN} is the fact that we are extrapolating, because the data that is going to be used after training is different from the data used to train it. Although looking similar to some training data, a pure Lorentzian curve does not exist in the simulated data, and a small difference between it and a pure Lorentzian spectrum can cause the \gls{ANN} to not generalize well, because \gls{ANN} are  used to interpolate functions and therefore, should not be applied to extrapolation  \cite{acharige2022machine}. There are several solutions to this problem. One of them consists in having a \gls{DL} model capable of extrapolating well. One such example are \glspl{PINN} that hard codes the physics of the problem into the \gls{ANN} and can extrapolate extremely well \cite{raissi2019physics}. The problem for our case is that we cannot extract simple differential equations that correlates the input to the output, and therefore, \glspl{PINN} cannot be considered. The last solution and the least efficient is to apply regularization to the training parameters. By doing so, we manipulate the bias-variance trade off, and by having a lower variance model with higher bias our \gls{ANN} is able to extrapolate with the downside of losing interpolation performance.

Matching the correct output given a spectrum becomes an arduous task, as it necessitates accurately aligning the corresponding parameters across all wavelengths encompassed within the spectrum. Training with only 810 data points is not enough to obtain reasonable results. A solution is to generate data with the surrogate model to be used as training for the \gls{MVNN}. We generated 7839 input-output pairs in \textcolor{black}{4 min and 13 seconds} on a Google colab's \gls{CPU}. The geometry limits are the same as for the simulations, the only difference is the step, which now is \SI{25}{nm} instead of \SI{65}{nm}. % The total time to generate the 7839 data was 7 min 31 s due to the time to generate the input matrices and save everything.

\begin{figure}[H]
\centering\includegraphics[width=0.85\textwidth]{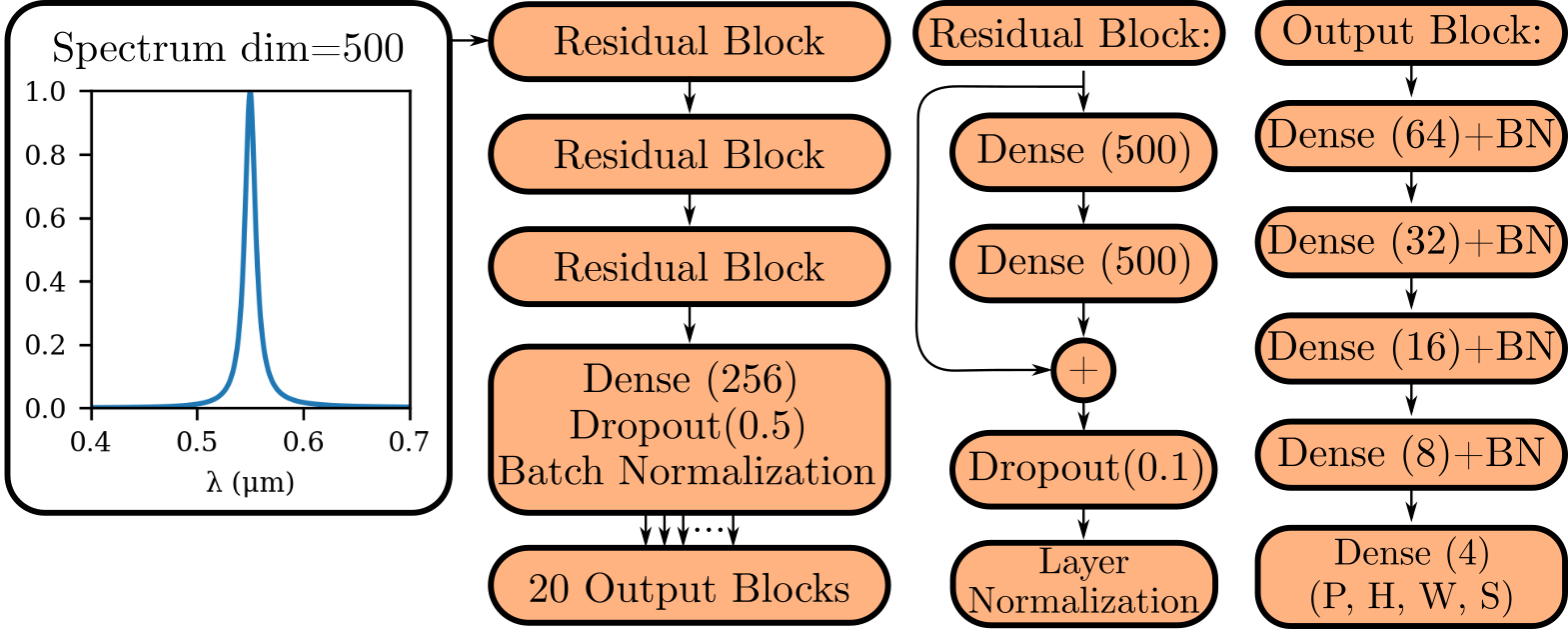}
\caption{Schematic of the \gls{MVNN} Architecture. We used 3 residual blocks with 500 neurons per layer followed by a layer with 256 neurons, a dropout layer with rate = 0.5 and batch normalization. Each output block consists of 5 layers, each followed by a batch normalization layer. The first layer starts with 64 neurons and geometrically decays to 4.} %Needs to be better
\label{fig:mvnn}
\end{figure}

The architecture of the \gls{MVNN} is represented by Figure \ref{fig:mvnn}. We opted to use residual blocks as they usually grant us good generalization with a reasonable depth \cite{he2020resnet, orhan2017skip}. On top of that, we also used $\mathrm{L_1}$ regularization, by summing all the weights on the network, and multiplying by the penalization strength, on this case, $10^{-7}$, as the \gls{ANN} had 2 023 808 trainable parameters, meaning an average penalization of 0.2 per parameter. All activation functions were \gls{GELU} except for the last layers of the output block and the kernel initialiser was Glorot normal for all parameters \cite{hendrycks2016gaussian, glorot2010understanding}. \gls{BN} Layers and dropout layers with high rate were employed to enhance the generalization and extrapolation capacity of the \gls{ANN}.
%%%%%%%%%%%%%%%%%%%%%%%%%%
%%%%%%%%%%%%%%%%%%%%%%%%
The model was trained with tensorflow for 100 epochs with patience of 10 \cite{abadi2016tensorflow}. However, differently from the surrogate model, we have not used holdout.

\textcolor{black}{The decision to exclude the use of holdout training for the \gls{MVNN} may be viewed as controversial, but it yielded superior initial estimates for the subsequent back-propagation optimization process. Identifying concrete reasons for this phenomenon can be challenging, as it involves various factors, such as the small size of the dataset used for training, the potentially inadequate number of outputs, or the inherent nature of the \gls{MVNN} in requiring a vast amount of training data for optimal performance in this particular scenario. Additionally, the notion of withholding data for assessing overfitting becomes less crucial in this context, as each additional data point could provide relevant information due to the high sensitivity of the dataset. It is important to note that the results obtained using the \gls{MVNN} are going to be used as initial guesses for back-propagation optimization and do not necessitate perfection.}

The \textcolor{black}{chosen} optimizer is Adam and the batch size has been fixed to 32 \cite{kingma2014adam}. The training stopped at epoch 100 with final loss of \textcolor{black}{$2.3093\times 10^{-4}$}. The total training time was \textcolor{black}{46 min 8 s on a google colab's \gls{CPU}}.

The loss function of the \gls{MVNN} illustrates a crucial characteristic of the behaviour of the model. This being said, that the presence of a single low loss output solution has a significant impact on the final loss value.
This implies that the attainment of a single high-quality solution effectively determines the calculated loss, thereby compelling the specialization of each individual output. Subsequently, we exploit the advantageous properties of the surrogate model \gls{ANN}, which is fully differentiable. This enables us to propagate gradients through the model using back-propagation, allowing for efficient refinement of the solutions initially provided by the \gls{MVNN}. For this specific study, we employed a configuration with 20 outputs, which proved to be sufficient in offering reasonable starting guesses for the back-propagation optimization. The loss function for the \gls{MVNN} is explained in the original material \cite{zhang2018multivalued}. A small modification of the loss function is used in this study and displayed on the supplementary information section S2.

To quantify the performance of the \gls{MVNN}, we calculated the loss function $L\left(\mathbf{w}\right)$  over a batch of size N, as represented by \Cref{eq:loss}.
%%%%%%%%%%%%%%%
\begin{equation}
    L\left(\mathbf{w}\right) = \frac{1}{N}\sum_{i=1}^N \frac{1}{B_i\left(\mathbf{w}\right)}
    \label{eq:loss}
\end{equation}
%%%%%%%%%%%%%%%%%%%%%%
The weights $\mathbf{w}$ are involved in the calculation of the loss, and $B_i\left(\mathbf{w}\right)$ represents the inverse of the loss for a single batch, defined as the sum of the inverse losses for each output, as:
%%%%%%%%%%%%%%%%%%%
\begin{equation}
    B_i\left(\mathbf{w}\right) = \sum_{j=1}^{N_{outs}}\frac{1}{S_j\left(\mathbf{w}\right)} 
\end{equation}
%%%%%%%%%%%%%%%%
Here $S_j\left(\mathbf{w}\right)$ is the $\mathrm{L_2}$ distance between the output $j$ of the \gls{MVNN}, $\mathbf{G}_{\mathrm{ANN},j}\left(\mathbf{x}, \mathbf{w}\right)$ given the input $\mathbf{x}$ and the desired response $\mathbf{d}$, 
%%%%%%%%%%
\begin{equation}
    S_j\left(\mathbf{w}\right) = ||\mathbf{G}_{\mathrm{ANN},j}\left(\mathbf{x}, \mathbf{w}\right) - \mathbf{d}||^2_2
    \label{eq:l2_mvann}
\end{equation}
%%%%%%%%%%
The motivation for the utilization of such loss function is to lower the losses for conflicting outputs and is explained in the original material \cite{zhang2018multivalued}. However, we give here a quick explanation If a single output matches very well, $S_j\left(\mathbf{w}\right) \rightarrow 0 \implies B_i\left(\mathbf{w}\right) \rightarrow \infty$ which will imply that the loss for that point of data is going to tend to 0 even if there is a $S_j\left(\mathbf{w}\right) \rightarrow \infty$ \cite{zhang2018multivalued}. The only difference between the original \gls{MVNN} and the used on this paper is that we haven't divided by 2 the $\mathrm{L_2}$ distance represented by \Cref{eq:l2_mvann} \cite{zhang2018multivalued}.

\begin{figure}[H]
\centering\includegraphics[width=0.8\textwidth]{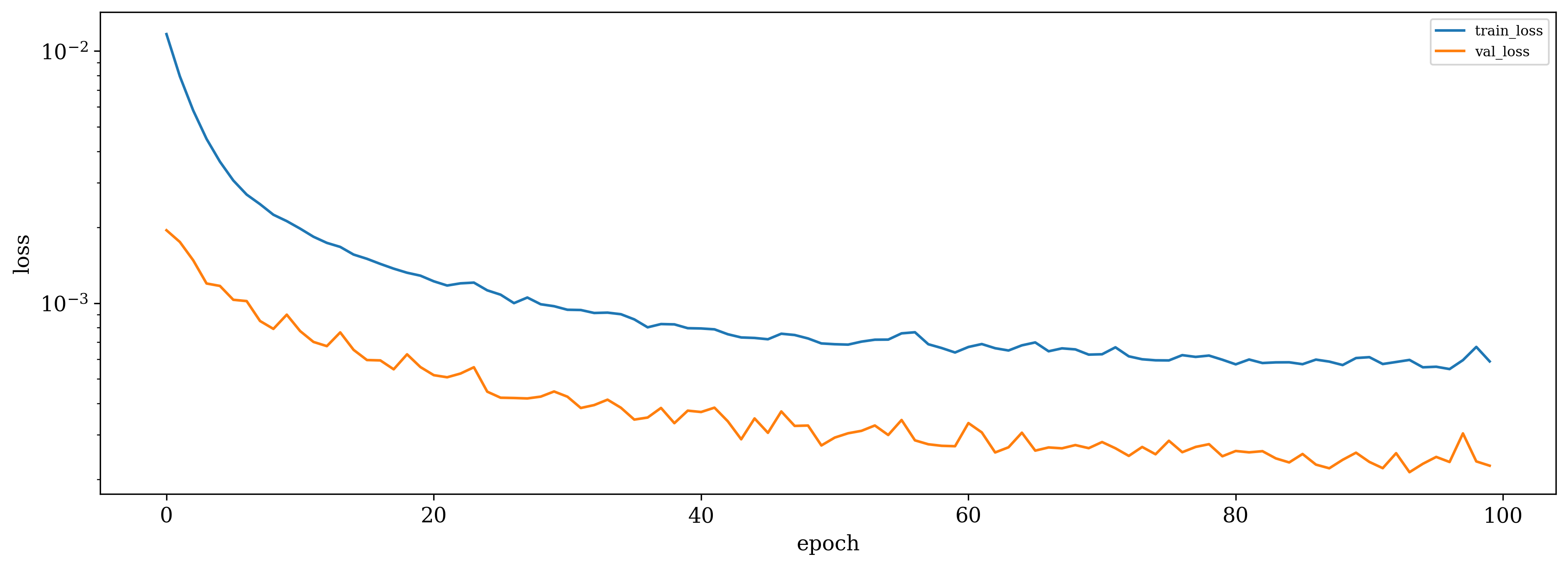}
\caption{\gls{MSE} loss of the training and validation sets in function of the epoch for the \gls{MVNN} model.
}
\label{fig:MVANN_training}
\end{figure}

Figure \ref{fig:MVANN_training} displays the loss evolution at each epoch. \textcolor{black}{The validation loss considers the same dataset used to train it, and it is lower than the train loss due to the use of dropout, batch normalization layers and most importantly, the L1 regularization term that is only present in the training loss and not in the validation.}

%%%%%%%%%%%%%%%%%%%%%%%%%%%%%%%%%%

\section{Study of the efficacy of the \gls{MVNN} with different dataset sizes} 

\begin{figure}[H]
\centering\includegraphics[width=\textwidth]{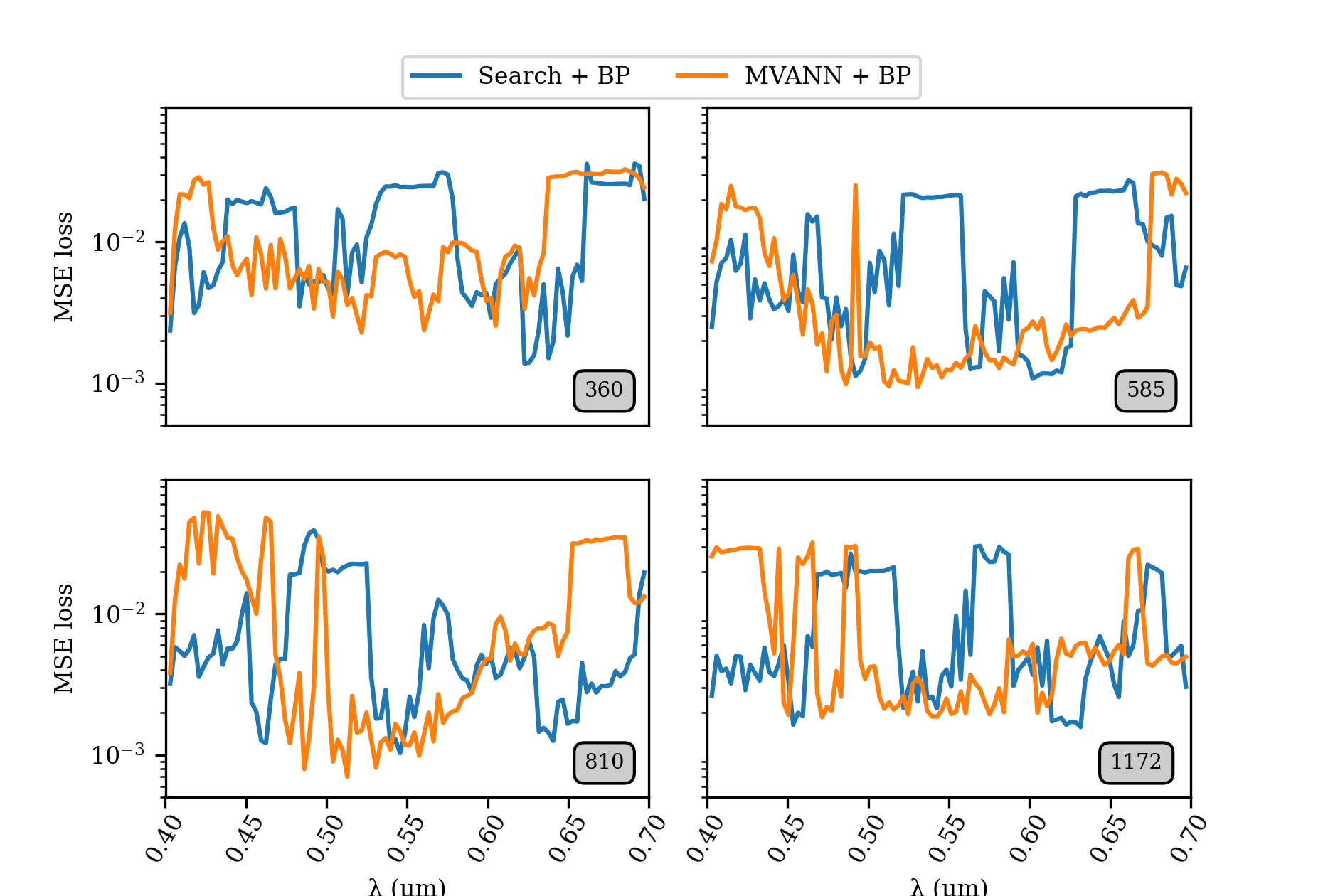}
\caption{ \gls{MSE} between target function and optimized design at each wavelength. The search scan with back-propagation  is represented by the orange curve, while and the \gls{MVNN} with back-propagation is given by the blue curve for 4 different dataset sizes.}
\label{fig:losses_cases}
\end{figure}

\textcolor{black}{In this section, we investigate how the dataset size affects the performance of the \gls{MVNN}. In \Cref{fig:losses_cases}, we present the \gls{MSE} between the target function and the optimized design at various wavelengths. The orange curve corresponds to the search with back-propagation, while the blue curve represents the \gls{MVNN} with back-propagation for four different dataset sizes. Notably, the \gls{MVNN} with back-propagation consistently outperforms the search method. The results from the 585 simulations demonstrate the lowest MSE for the \gls{MVNN} with back-propagation. This case, involving 585 simulations, essentially represents the limit beyond which further dataset size increases do not yield performance improvements. In other words, increasing the dataset size beyond this point does not enhance performance. Also, it can be observed that it is more common for the \gls{MVNN} to outperform the search.}
%%%%%%%%%%%%%%%%%%%%%%%%%%%%%%%%
%%%%%%%%%%%%%%%%%
%%%%%%%%%%%%%%%%%%%%%%%%%%%%%%%%
\section{Comparison between \gls{MVNN} and state-of-the-art algorithms}
\begin{figure}[H]
\centering\includegraphics[width=\textwidth]{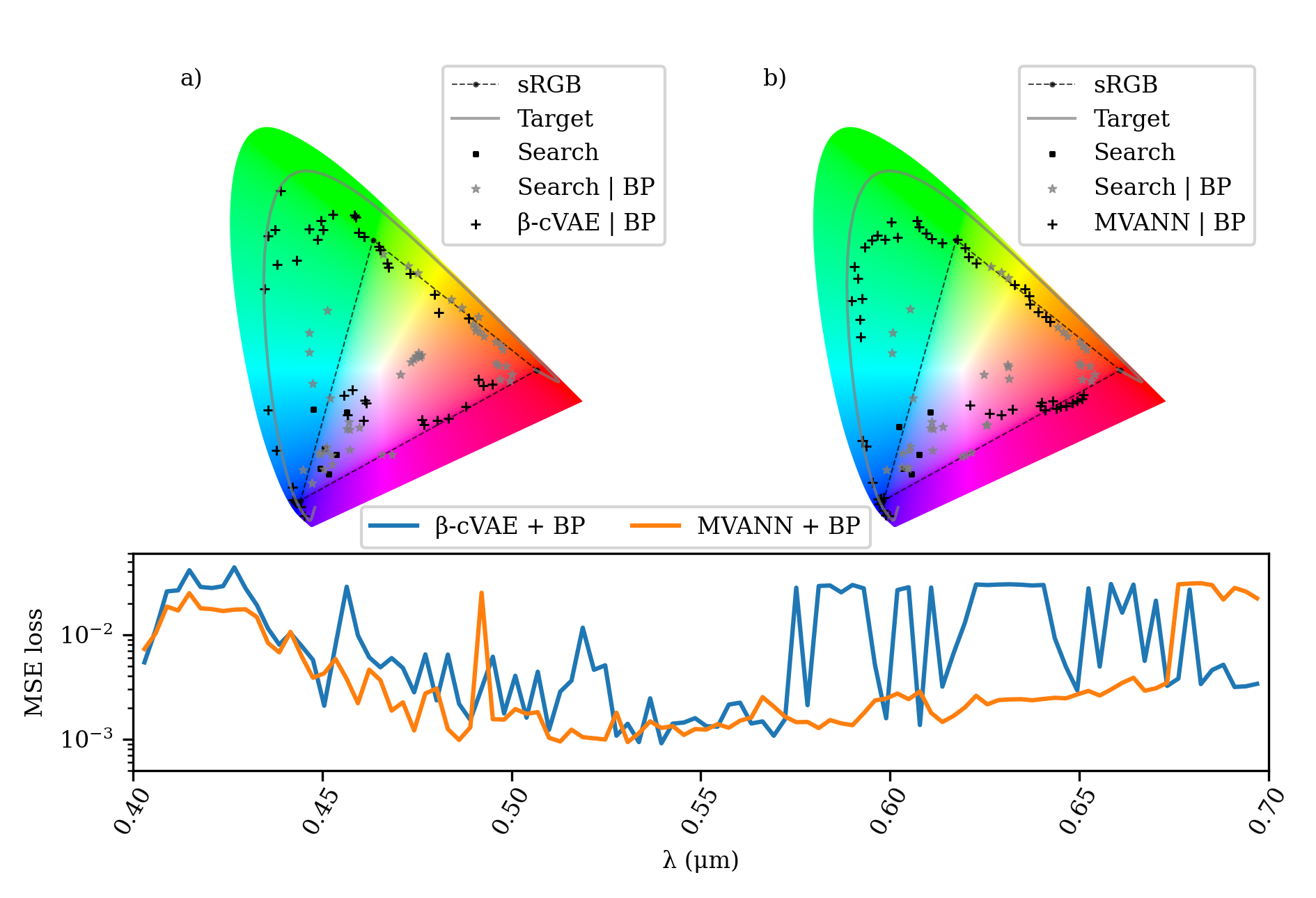}
\caption{Chromaticity diagram and \gls{MSE} loss comparing the target with the simulated spectrum of the optimized design for the a)  \gls{cVAE} and b) \gls{MVNN} after back-propagation for a dataset of size 585. They exhibit similar performance in terms of MSE and the chromaticity diagram, except for targets above 575 nm.}
\label{fig:cVAE_585_comp}
\end{figure}

% compare MVANN and CVAE,
\textcolor{black}{In \Cref{fig:cVAE_585_comp}, we compared the \gls{MVNN} with a \gls{cVAE} \cite{lin2022machine, yu2022inverse}. The \gls{cVAE} model utilized a similar architecture to the \gls{MVNN}, comprising 3 645 288 trainable parameters and having 2 latent dimensions. Training was conducted over 100 epochs using AdamW with a learning rate of $3\times 10^{-4}$ and \gls{MSE} loss. The training process took 41 minutes and 40 seconds. For comparison, training the \gls{MVNN} on the same CPU typically requires around 50 minutes. The results can be seen in \Cref{fig:cVAE_585_comp}.
One of the challenges associated with the \gls{cVAE} is that it necessitates an optimization technique for determining the latent variables, unlike the \gls{MVNN} where the solution is typically present in at least one of the outputs. In our case, we applied a brute-force optimization by randomly guessing 20 values to make the comparison with the \gls{MVNN} more equitable. Furthermore, similar to the \gls{MVNN}, we employed back-propagation optimization for the design variables.
The \gls{cVAE} encoder consists of a first layer with 500 neurons, followed by 3 residual layers, and then 128 neurons. The decoder follows the same architecture, with the addition of a layer with 4 neurons after the one with 128 neurons. All layers utilize \gls{GELU} activation with Glorot normal initialization, except for the layers responsible for computing the mean and the logarithm of the variance before the sampling layer.
In conclusion, as demonstrated in \Cref{fig:cVAE_585_comp}, both methods generally exhibit similar performance. However, the \gls{cVAE} has two notable drawbacks when compared to the \gls{MVNN}. Firstly, it still requires an optimization technique to find the latent variables, while in the \gls{MVNN}, the solution is typically present in at least one of the outputs. Secondly, the implementation of the \gls{cVAE} is less straightforward and more convoluted compared to the \gls{MVNN}.}
% %%%%%%%%%%%%%%%%%%%%%%%%%%%
%%%%%%%%%%%%%%%%%%%%%%%%%%%
%%%%%%%%%%%%%%%%%%%%%%%%%%%%%%%%%%%%%%%
\textcolor{black}{One metric to benchmark the \gls{MVNN} and the Search after back-propagation is to compare the average \gls{MSE} and standard deviation. However, it does not tell the whole story, as it is observed a bi-stable regime, where it oscillates between two values, and therefore, the minimum and maximum values are also taken into account. \Cref{tab:comp} benchmarks the different dataset sizes.}
%%%%%%%%%%%%%%%%%%%%%%
\begin{table}[H]
\centering
\caption{Table representing the mean, standard deviation ($\sigma$), minimum and maximum values for the \gls{MSE} loss between the simulated and target spectrum for different dataset sizes and methods. The values are multiplied by $10^3$.}
\begin{tabular}{|cc|c|c|c|c|}
\hline
\multicolumn{2}{|c|}{Values in $10^{-3}$} & Mean  & $\sigma$ & min  & max   \\ \hline
\multicolumn{1}{|c|}{\multirow{2}{*}{360}}     & \gls{MVNN} + BP     & 12.8  & 10.37  & 2.28 & 32.82 \\
\multicolumn{1}{|c|}{}                         & Search + BP    & 13.92  & 9.8  & 1.37 & 35.96 \\ \hline
\multicolumn{1}{|c|}{\multirow{3}{*}{585}}     & \gls{MVNN} + BP     & 6.19  & 8.26  & 0.94 & 31.16 \\
\multicolumn{1}{|c|}{}                         & Search + BP    & 9.65  & 8.23  & 1.07 & 27.40 \\ 
\multicolumn{1}{|c|}{}                         & \gls{cVAE} + BP    & 12.06  & 11.98  & 0.91 & 44.14 \\
\hline

\multicolumn{1}{|c|}{\multirow{3}{*}{810}}     & \gls{MVNN} + BP     & 13.78 & 15.15 & 0.70 & 52.74 \\
\multicolumn{1}{|c|}{}                         & Search + BP    & 7.94  & 8.42  & 1.03 & 39.30 \\
\multicolumn{1}{|c|}{}                         & \gls{cVAE} + BP & 16.1 & 15.51 & 0.81 & 56.86\\ \hline
\multicolumn{1}{|c|}{\multirow{2}{*}{1172}}    & \gls{MVNN} + BP     & 9.37  & 10.26 & 1.86 & 32.15 \\
\multicolumn{1}{|c|}{}                         & Search + BP    & 9.55  & 8.53  & 1.58 & 30.37 \\ \hline
\end{tabular}

\label{tab:comp}
\end{table}
%%%%%%%%%%%
\textcolor{black}{It can be seen that the best case scenario is the dataset with 585 simulations, showing the \gls{MVNN} after back-propagation to have the least mean, however, for 810, the minimum is lower, and considering \Cref{fig:losses_cases}, that minimum lasts from \SI{475}{nm} to \SI{575}{nm}, which is mostly the whole green and cyan areas. However, this region exhibits instability and spikes.
Other conclusion is the whole stability of the \gls{MVNN} with back-propagation in the scenario with 585, outperforming the search in multiple areas.}
%%%%%%%%%%%%%

\section{Time table}

\begin{table}[H]
\centering
\caption{Time required for various stages, including data generation, model training, and optimization runs. The simulations were conducted on the Google Colab platform equipped with an Intel® Xeon® E5 v4 Processor operating at 2.2 GHz. It should be noted that determining the optimization time per design in the fixed height case can be misleading, as the back-propagation time does not scale linearly with the number of optimizations due to training on the same batch. Furthermore, the time needed to identify designs with the closest heights exhibits a significant increase as the number of simultaneous optimizations rises. For instance, while 6 designs required a mere 0.8 seconds, this duration is extended to 5 minutes and 30 seconds when dealing with 10 designs.}
\begin{tabular}{ll|ll}
\multicolumn{2}{l|}{Simulations}                                                                   & \multicolumn{2}{l}{Models Training}                                                                 \\ \hline
Total                                              & 33 min 40 s                                  & Surrogate                                           & 9 min 48 s                                    \\
Average                                               & 3.4 s                                         & Generation                                          & 4 min 13 s                                    \\
                                                   &                                               & MVANN                                               & 46 min 8 s                                   \\
                                                   &                                               & Total                                               & 1h 0 min 9 s                                   \\ \hline
\multicolumn{2}{l|}{\begin{tabular}[c]{@{}l@{}}Free height optimization\\ (1 design)\end{tabular}} & \multicolumn{2}{l}{\begin{tabular}[c]{@{}l@{}}Fixed height optimization\\ (6 designs)\end{tabular}} \\ \hline
Simulations                                        & 51.8 s                                          & Simulations                                         & 2 min 57 s                                     \\
Back-propagation                                   & 37.3 s                                          & Back-propagation                                    & 1 min 18 s                                    \\
Predictions                                        & 5.1 s                                         & Predictions                                         & 15 s                                          \\
Search                                             & 2.5 ms                                        & Total                                               & 4 min 30 s                                    \\
Total                                              & 1 min 34 s                                    & Per design total                                    & 35 s                                   
\end{tabular}
\label{tab:times}
\end{table}

% Bibliography
\bibliographystyle{unsrt}
\bibliography{references}